\documentclass[nofootinbib,aps,superscriptaddress,11pt]{revtex4-1}
\usepackage{geometry}
\usepackage{tabularx}
\usepackage{graphicx, subfigure}
\usepackage{dcolumn}
\usepackage{bm}
\usepackage{braket}
\usepackage{epsfig,amsmath}
\usepackage{amssymb}
\usepackage{subfigure,esint}
\usepackage{float}
\usepackage[usenames,dvipsnames]{color}
\usepackage[pagebackref=false, colorlinks=true]{hyperref}
\definecolor{redish}{rgb}{0.7,0.2,0.0}  
\definecolor{bluish}{rgb}{0.2,0.5,0.8}

\hypersetup{linkcolor=redish,          
	citecolor=blue,        
	filecolor=magenta,      
	urlcolor=blue}          

\begin{document}

\author{Mamta Gautam}\email{mamtag@iitk.ac.in}
\affiliation{Department of Physics, Indian Institute of Technology Kanpur, \\ Kanpur 208016, India}
\author{Nitesh Jaiswal}\email{nitesh@iitk.ac.in}
\affiliation{Department of Physics, Indian Institute of Technology Kanpur, \\ Kanpur 208016, India}
\author{Ankit Gill}\email{ankitgill20@iitk.ac.in}
\affiliation{Department of Physics, Indian Institute of Technology Kanpur, \\ Kanpur 208016, India}
\title{Spread Complexity in free fermion models }

\begin{abstract}
We study spread complexity and the statistics of work done
for quenches in the three-spin interacting Ising model, the
XY spin chain, and the Su-Schrieffer-Heeger model.
We study these models without quench and for different
schemes of quenches such as sudden quench and multiple sudden
quenches. We employ the Floquet operator
technique to investigate all three models in the presence of
time-dependent periodic driving of parameters. In contrast to
the sudden quenched cases, the periodically varying
parameter case clearly shows non-analytical behaviour near the critical
point. We also elucidate the relation between work done and
the Lanczos coefficient and how the statistics of work done
behave near critical points.
\end{abstract}

\maketitle

\section{Introduction}
\label{intro}
The notion of complexity for a quantum state has become a well-known probe of interesting physical phenomena like quantum chaos and quantum phase transitions \cite{sachdev}. 
There are several variants of complexity currently available in the literature \cite{Chapman:2021jbh}. Perhaps the most studied example is the 
Nielsen complexity, which measures the least number of unitary operations to construct a target state
from a given reference state \cite{myers, khan, praj, gautam, Bhattacharyya:2018bbv, Jaiswal:2020snm}. More recently, Balasubramanian et al. \cite{balas} came out with the concept of \textit{spread complexity} (SC), which will
be the focus of this work. 

Recall that in a geometrical framework, the computation of Nielsen complexity reduces to finding the geodesic 
connecting the target and reference states. For this, we require to make specific choices of the unitary gates, their penalties, and cost functionals. 
These requirements might raise legitimate concerns as to which features of the quantum system the complexity actually depends on. 
Therefore, one has to look for a more general form of complexity, which was addressed recently in \cite{balas}, where the concept of SC of a quantum state under a unitary evolution was introduced. Roughly, the SC
measures the spread of wave function over some fixed basis. Suppose that we start with an initial state and allow it to spread with time over a given basis; then, one can define SC by minimizing the cost associated with spread over all possible basis. Importantly, it was proved in \cite{balas} that the basis set which minimises the cost associated with the wavefunction spreading is nothing but the Krylov basis, which can be constructed by using the standard Gram-Schmidt orthogonalization process starting from the initial state \cite{parker, Viswanath, Lanczos}.  
Unlike Nielsen complexity, SC requires only the 
reference and target states, given the Hamiltonian of the system. The algorithm for its construction is also computationally efficient. Given these advantages, SC has therefore received the attention of late \cite{gcaptua, pratik, kcaptua, parker, Lanczos, shir, captua, Nizami,  sayantanb, sayantan}. Among these works,  \cite{captua}
computed SC for the Su-Schrieffer-Heeger (SSH) model and found that, in a quench protocol, SC can distinguish the different topological phases of the SSH system. This result was also further supported in a later work \cite{kcaptua}, where the SC was used to differentiate the two topological phases of a Kitaev chain, which establishes the SC as a standard probe of the quantum phase transition.

In line with the above discussion, in this paper, we are interested in studying the nature of SC of non-equilibrium dynamics in many-body quantum systems. 
Recall that a system can be taken out of the equilibrium in several ways: by applying a driving field 
or by pumping energy and particles into the system, etc. Here we will study the behaviour of SC in such non-equilibrium dynamics by changing one of 
the parameters of the system, i.e., by employing a  quantum quench \cite{quan, venuti, polko}. In this paper, we calculate the SC for three paradigmatic models: the 
transverse XY spin chain, the Ising model with three-spin interactions, and the Su-Schrieffer-Heeger model (SSH), where the evolution is assumed to be under 
both time-dependent and time-independent Hamiltonians. We start with a brief discussion on the construction of the Krylov basis and the procedure to calculate the associated Lanczos coefficients via the geometric method, first introduced in \cite{gcaptua}. Then we explain the dynamics of SC for a single sudden quench, multiple sudden quenches, and periodic driving for all three spin chains. In particular, in a time-dependent scenario, we calculate the SC by periodically changing the parameter over the stroboscopic time.

Finally, recalling that a quantum quench is analogous to a classical thermodynamic 
transformation and is characterised by usual  thermodynamic quantities, such as the work done on the system, entropy produced, and heat exchanged \cite{ kurchan, talkner}, here we compute the average and the variance of the work done on the 
system for a quantum quench \cite{silva}. We also discuss the relation
between the Lanczos coefficients and the average and the variance of the distribution of the work done in these systems.

Let us start with a brief discussion of the construction of the Krylov basis, which plays a main role in computation and the subsequent definition of 
 SC following \cite{balas,gcaptua,kcaptua}.
Consider a general quantum state $\ket{\psi(s)}$, related to another one  by the unitary transformation
$\ket{\psi(s)}=\exp(-i H s)\ket{\psi(0)}$.
We call $\ket{\psi(0)}$ as the reference state and $\ket{\psi (s=1)}$ as the target state. 
The parameter $s$ is the circuit parameter that 
connects the reference and target states, running from $0$ to $1$. To calculate the SC, the target state is expanded in a given 
basis $\{\ket {\psi_{n}}\}$
\begin{equation}
\ket{\psi(s)}=\sum_{n=0}^{\infty}\frac{(is)^{n}}{n!}\ket{\psi_{n}}~,
\end{equation}
where $\ket{\psi_{n}}=H^{n}\ket{\psi(0)}$, $H$ being the Hamiltonian
generating the above unitary transformation. The Gram-Schmidt procedure is applied to these states to obtain a
orthonormal basis ($\mathcal{K}=[\ket{K_{n}}$,$n$=0,1,2....]). These basis vectors form a subspace of the Hilbert space, 
and have a dimension generally less than that of the Hilbert space. To quantify how much $\ket{\psi(0)}$ 
spreads over the Hilbert space while evolving 
under a  given Hamiltonian, we consider the following  cost function for an arbitrary orthonormal basis $\mathcal{B}=\ket{B_{n}}$
\begin{equation}\label{cost}
\mathcal{C_{\mathcal{B}}}(s)=\sum_{n}n|\bra{\psi(s)}\ket{B_{n}}|^{2}~,
\end{equation}
and then subsequently minimise it over all choices of basis. 
Spread complexity is then defined as
\begin{equation}
\mathcal{C}(s)=\textit{min}_{\mathcal{B}}\mathcal{C_{\mathcal{B}}}(s)
\label{sprkry}~.
\end{equation}
The Hamiltonian in the Krylov basis is tridiagonal, i.e.,
\begin{equation}
H\ket{K_{n}}=a_{n}\ket{K_{n}}+b_{n}\ket{K_{n-1}}+b_{n+1}\ket{K_{n+1}}~,
\end{equation}
where the coefficients $a_{n},b_{n},b_{n+1}$ are known as the Lanczos coefficients.
Similarly, the evolved state can be expanded in terms of the Krylov basis as 
\begin{equation}
\ket{\psi(s)}=\sum_{n}\psi_{n}(s)\ket{K_{n}}~,
\label{Krylovstat}
\end{equation}
and it can be checked that the evolved state  satisfies the following discrete Schrodinger equation
\begin{equation}
    i\partial_{s}\psi_{n}(s)=a_{n}\psi_{n}(s
)+b_{n}\psi_{n-1}+b_{n+1}\psi_{n+1}
\end{equation}
It was proved in \cite{balas}, when evaluated in  the Krylov basis, the cost defined in Eq. \eqref{cost} becomes minimum,
 and hence the SC is defined as 
\begin{equation}
\mathcal{C}(s)=\sum_{n}n|\psi_{n}(s)|^{2}\label{spreadkry}~.
\end{equation}
With this brief introduction, we now move to the main part of the paper and consider SC in different free fermion models. 

\section{Three spins interacting transverse Ising model}\label{3spin}
The Hamiltonian of the transverse Ising model with three spin interactions is given by \cite{Kopp,threediva}:
\begin{equation}
\mathcal{H}=-\frac{1}{2}\Big(\sum_{i}\sigma^{z}_{i}\Big[h+J_3\sigma^{x}_{i-1}\sigma^{x}_{i+1}\Big]
+J_{x}\sum_{i}\sigma^{x}_{i}\sigma^{x}_{i+1}\Big)\label{three hamilt}~,
 \end{equation}
Where $\sigma^{z}$ and $\sigma^{x}$ are Pauli matrices, $J_{x}$ is the strength of the nearest neighbour ferromagnetic interaction, 
and $J_3$ is the strength of three spin interactions. In the limit $J_3 \to 0$, this model reduces to the standard transverse Ising model. 
Here, $J_{x}$ will be set to unity without loss of generality. Even in the presence of the three-spin interaction term, the Hamiltonian 
given by Eq. (\ref{three hamilt}) is exactly solvable \cite{threediva}. By a Jordan-Wigner(JW) transformation \cite{Jw1, Jw2, Jw3} and a Fourier transforms, 
we can map the Hamiltonian (\ref{three hamilt}) into a free fermion model \cite{threediva}. 
\begin{equation}
H=-\sum_{k>0}(h+\cos{k}-J_3\cos{2k})(c^{\dagger}_{k}c_{k}+c^{\dagger}_{-k}c_{-k})+i(\sin{k}-J_3\sin{2k})(c^{\dagger}_{k}c^{\dagger}_{-k}+c_{k}c_{-k})\label{three fourier}~.
\end{equation}
The Hamiltonian in Eq. (\ref{three fourier}) can be further diagonalized by using the Bogoliubov transformation, which yields the excitation spectrum 
\begin{equation}  
  \epsilon_{k}=\sqrt{h^{2}+1+J_3^{2}+2 h\cos{k}-2h J_3\cos{2k}-2J_3\cos{k}}=|\Vec{R}|\label{three gap}.
\end{equation}
From Eq. (\ref{three gap}), it is seen that the energy gap vanishes at $h=J_3+1$ and $h=J_3-1$ for $k=\pi$, and $k=0$ modes, respectively. 
These two lines correspond to quantum phase transitions (QPTs) from paramagnetic to ferromagnetic ordered phases. 
The wave vector at which a minimum of $\epsilon_{k}$ occurs shifts from $k=0$ to $k=\pi$ when one crosses the line $h=J_3$.
There is an additional phase transition at $h=-J_3$; this is an anisotropic transition. 
By taking $\frac{\partial\epsilon_{k}}{\partial k}=0$, we find out the values of wave vector $k$ at which Eq. (\ref{three gap}) 
is minimum. Beside $k=0$ and $k=\pi$, we have 
\[
\cos{k}=\frac{h-J_3}{4hJ_3}~,
\]
At which the energy gap is minimum. By using the above value of $k$ in Eq. (\ref{three gap}), one can see that $\epsilon_{k}$ is gapless at 
$h=-J_3$, and this line will intersect the $k=0$ line at $J_3=0.5$. Therefore, for $J_3<0.5$, the anisotropic transition cannot occur. 
\\
To study SC, we need to analyse the ground state and write it in terms  of the Krylov basis.
The Hamiltonian and ground state can be written in the following form (see Appendix \ref{3spinapp})
\begin{equation}
    H=-\sum_{k>0} I R\cos\phi -\sum_{k>0}\Big[R\cos\phi J^{k}_{0}-i R\sin\phi (J^{k}_{+}-J^{k}_{-})\Big]\label{SU2_three2}~,
\end{equation}
where
\begin{equation}
\cos{\phi}=\frac{R_3}{R}~,\qquad
R_3=h+\cos{k}-J_3\cos{2k}~,\qquad R_2=J_3\sin{2k}-\sin{k}~,\qquad R=\sqrt{R_2^{2}+R_3^{2}}=|\vec{R}(k)|\label{para3}~.
\end{equation}
Now it is easy to see that $J_{0}^{k}$, $J_{+}^{k}$, $J_{-}^{k}$, defined in Eq. 
\eqref{su2_generators} satisfy the commutation
relations of  $su(2)$ algebra, and hence, the Hamiltonian in Eq. \eqref{SU2_three2} is actually an element 
of this algebra.

The ground state for a single momentum mode can be  written as
\begin{equation}
\ket{\psi_k(s)}=e^{-i\frac{s \phi}{2}(J_{+}^{k}+J_{-}^{k})}\ket{\frac{1}{2},-\frac{1}{2}}_{k}
\label{circuitst}
\end{equation}
where $s=1$ corresponds to the target state. We now want to calculate the SC of creating this state starting from
a suitable reference state.
Here as the  reference state, we make the most straightforward choice, 
i.e. fermion vacuum $\ket{0,0}$ which can be represented as $\ket{\frac{1}{2},-\frac{1}{2}}_k$.
First, we write down the ground state in the following suitable form
\begin{equation}
\ket{\psi_{gs}}=\prod_{k>0}\cos{\frac{\pi-\phi}{2}}\exp\Big[e^{-i\frac{\pi}{2}}\tan{\frac{\pi-\phi}{2}}J_{+}^{k}\Big]
\ket{j,-j}_{k}~,
\end{equation}
or
\begin{equation}
\ket{\psi_{gs}}=\prod_{k>0}\exp\Big[\frac{\pi-\phi}{2} e^{-i\frac{\pi}{2}}J_{+}^{k}-\frac{\pi-\phi}{2}e^{i\frac{\pi}{2}}J_{-}^{k}\Big]\ket{j,-j}_{k}~,
\end{equation}
where, in this case $j=\frac{1}{2}$.

Since the target state is a $SU(2)$ coherent state, the Krylov basis vectors are just \cite{gcaptua, balas}
\begin{equation}
\ket{K_{n}}=\ket{-j,j+n}=\sqrt{\frac{(2j-n)!}{n!(2j)!}}J_{+}^{n}\ket{j,-j}\label{Krylov}~.
\end{equation}
Following a procedure similar to that of the one outlined in \cite{captua}, we calculate the 
 the SC for  single momentum mode is to be 
\begin{equation}
\mathcal{C}_k(s=1)=\left(\cos{\frac{\phi}{2}}\right)^{2}  ~.
\end{equation}
Taking the continuum limit, we obtain the SC of the ground state as 
\begin{equation}
\mathcal{C}=\frac{1}{2\pi}\int_{0}^{\pi}dk \left(\cos{\frac{\phi}{2}}\right)^{2} ~.
\label{com three}
\end{equation}
In the continuum limit, we will work with the quantities per system size 
, so we suppress the factor $N$. And follow this notation throughout the paper. SC acts as a tool to probe quantum phase transitions has been discussed in \cite{captua,kcaptua}, via Eq. (\ref{com three}).
Here, we take the derivative of Eq. (\ref{com three}) with respect to $h$ and plot the same in Fig. \ref{fig: derivative com}, and find that
this shows a peak around critical points $J_3+1$ and $J_3-1$. There are three peaks for $J_3>0.5$, one of which 
corresponds to anisotropic transition $h=-J_3$.

\subsection{Spread complexity evolution under a single quench}
We now study the time dependence of the SC under a single sudden quench. 
We consider  unitary time evolution of an initial state $\ket{\psi_{i}}$ 
(preferably eigenstate of an initial Hamiltonian $H_{i}$) under  a different Hamiltonian $H_{f} (h_{f},J_{3f})$,
which has parameter values different from the  initial one.
The evolved state  at an arbitrary time  is given by 
$\ket{\psi(t)}=\exp(-i H_{f} t)\ket{\psi_{i}}$, 
where we take $\ket{\psi_{i}}$ to be  the ground state of initial Hamiltonian $H_{i}$ with parameters $(h_{i},J_{3i})$.

\begin{figure}[ht!]
 \centering
 \subfigure[]{
 \includegraphics[width=0.3\textwidth]{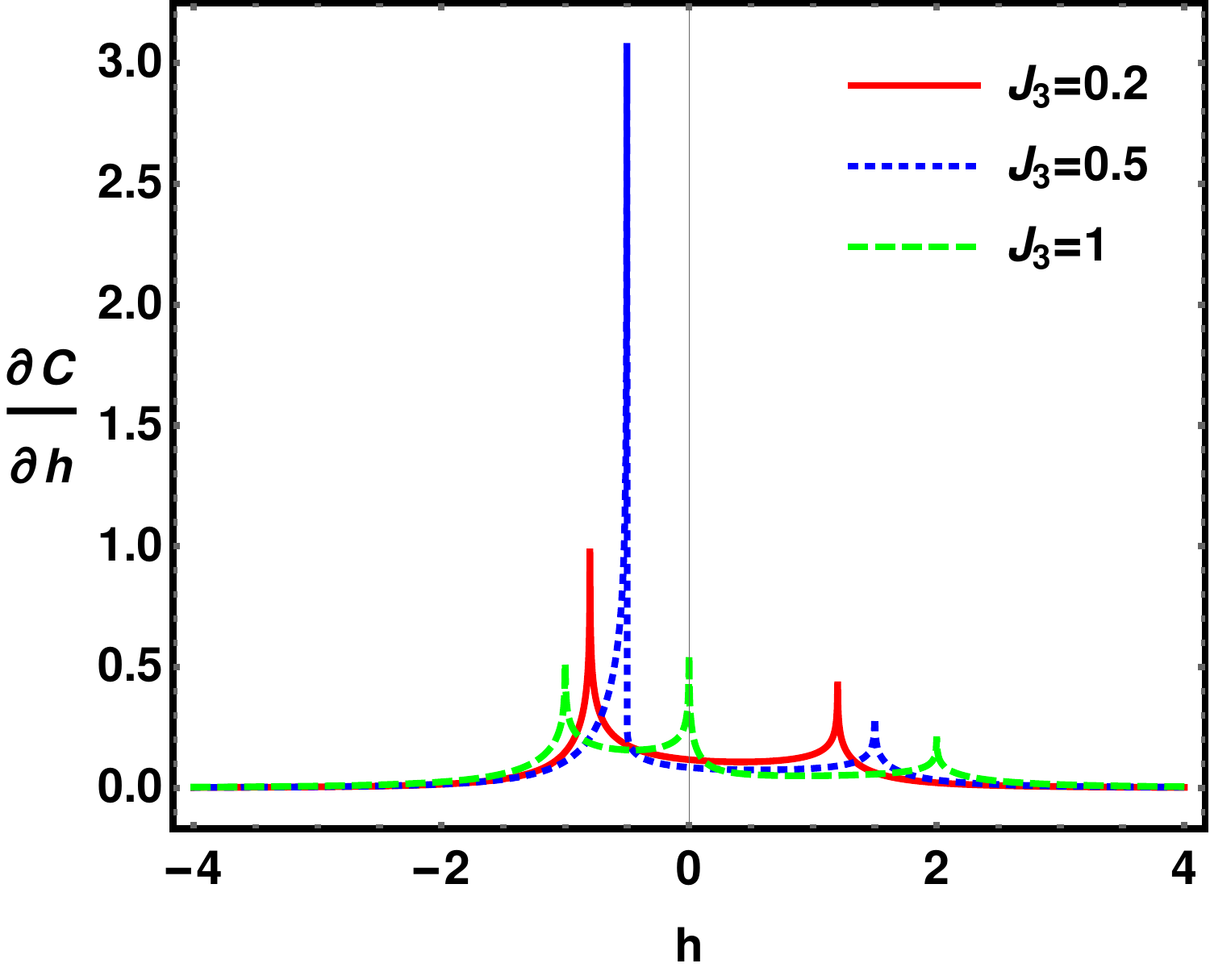}
\label{fig: derivative com}}
 \subfigure[]{
 \includegraphics[width=0.3\textwidth]{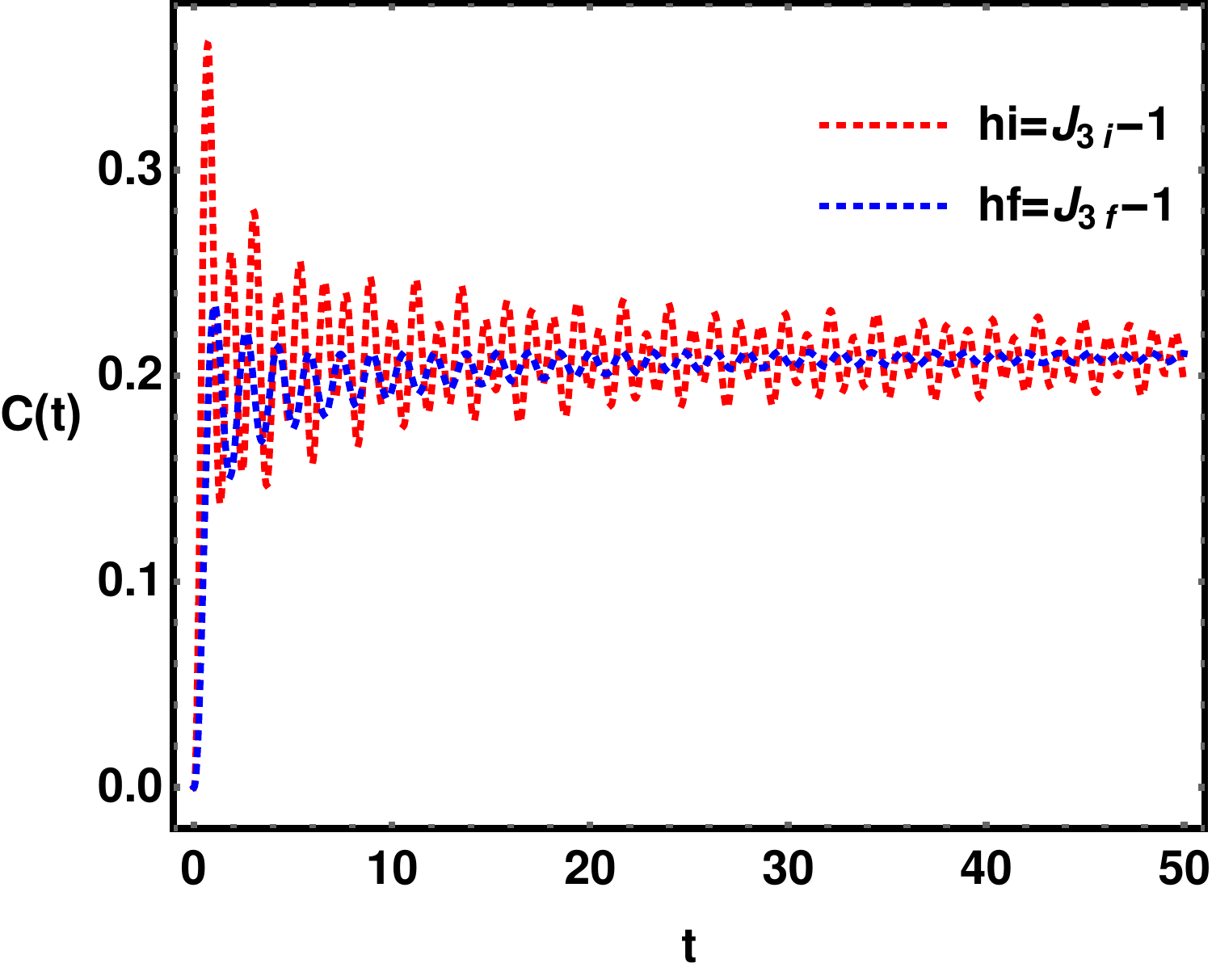}
 \label{fig:squench} }
  \subfigure[]{
 \includegraphics[width=0.3\textwidth]{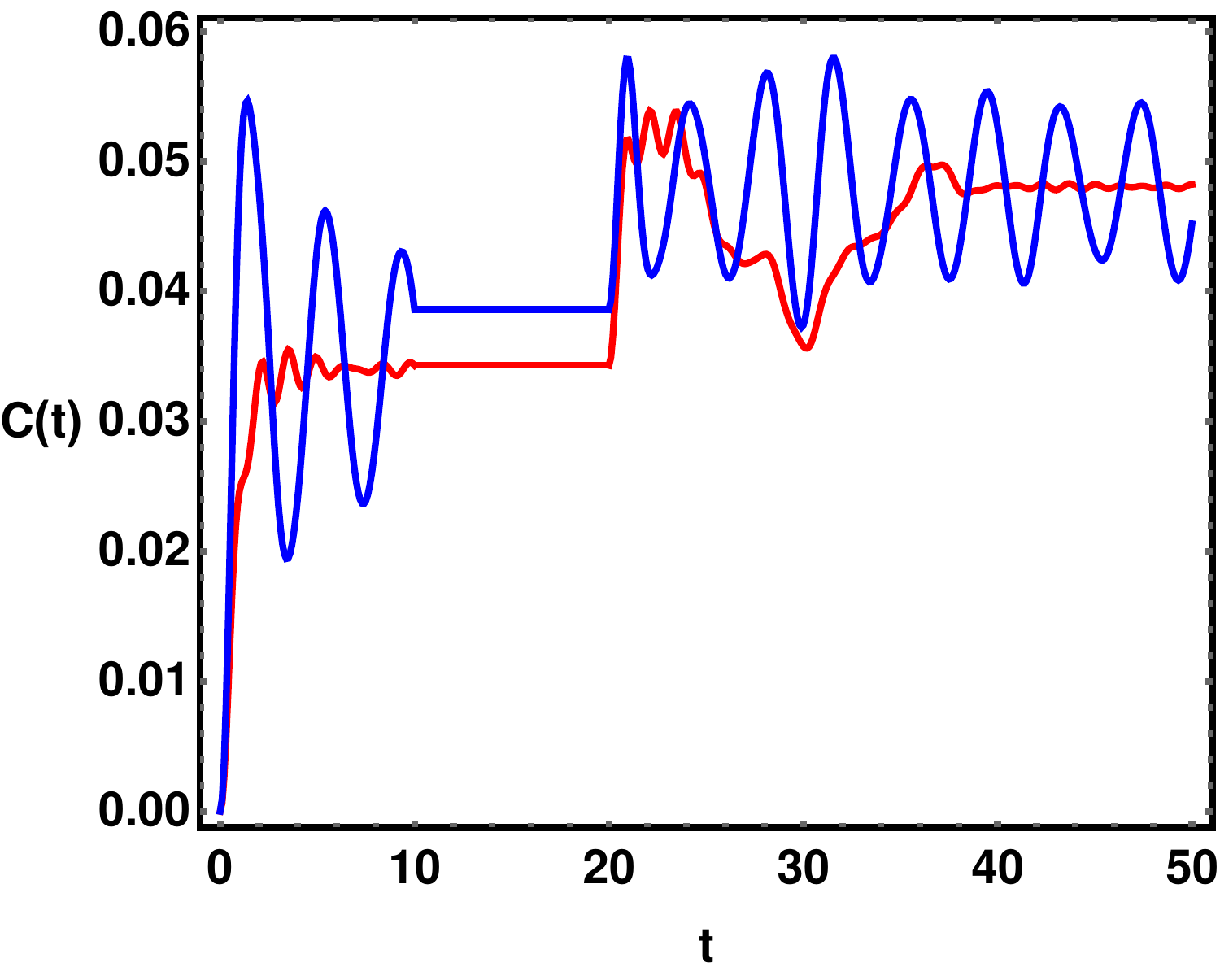}
 \label{fig:multi} }
  \caption{\small{SC of the transverse Ising model with three spin interactions : (a)\, Derivative of complexity with respect to $h$, (b)\, SC evolution after a single quench, and (c)\, SC after  multiple quenches.}}
 \end{figure}

We first find out  the return amplitude defined  as $S=\braket{\psi(t)|\psi_{i}}$. Since, $\ket{\psi_{i}}=\mathcal{D}(z_{i})\ket{j,-j}$,
with 
\begin{equation}
\mathcal{D}(z_{i})=\exp\Big[\frac{\pi-\phi_{i}}{2}e^{-i\frac{\pi}{2}}J_{+}-\frac{\pi-\phi_{i}}{2}e^{i\frac{\pi}{2}}J_{-}
\Big]~,
\label{grndkry}
\end{equation}
 this is given by
\begin{equation}
S=\braket{\psi(t)|\psi_{i}}=\bra{j,-j}\mathcal D^{\dagger}(z_{i})\exp{(i H_{f}t)}\mathcal{D}(z_{i})\ket{j,-j}~.
\label{returnamp}
\end{equation}
By using  the Baker-Campbell-Hausdorff (BCH) formula, we can write down the analytical expression for the return 
amplitude to be 
\begin{equation}
S=\cos(R_{f} t)-i\cos(\phi_{f}-\phi_{i})\sin(R_{f} t)~.
\end{equation}
SC,  in continuum limit is given by  (see \cite{captua}, and Appendix \ref{3spinapp}) 
\begin{equation}
\mathcal{C}(t)=\frac{1}{2\pi}\int_{0}^{\pi}\Big(1-|S|^{2}\Big)dk=
\frac{1}{2\pi}\int_{0}^{\pi}\sin^{2}(\phi_{f}-\phi_{i})\sin^{2}(R_{f}t)dk \label{s1quench}~.
\end{equation}
We plot the SC (Eq. (\ref{s1quench})) in Fig. \ref{fig:squench2xy},
where the red curve denotes the evolution when the initial value of parameters are at critical points, 
and the blue curve is when the final parameters are at critical points. The numerical values of the parameters for the red curve: 
$J_{3i}=0.4$, $J_{3f}=1$, $h_f=1.5$, and  blue curve: $J_{3i}=1$, $J_{3f}=0.4$, $h_i=1.5$.
From these plots, we see that at the  late times, $\sin{(R_{f} t)}$ attain a constant value. 
It is clearly seen from Fig. \ref{fig:squench} that there is no sign of QPT in the evolution of the SC,
even when we take the parameters at critical points.

\subsection{Spread complexity evolution after multiple quenches}
In a multiple-quench scenario, we  consider the succession of three quenches.
\begin{itemize}
\item  We prepare the state of $H_{i} (h_{i}, 
 J_{3i})$ at $t=0$ and evolve it for $10$ using $U_{1}(t)=\exp[-i H_{f} (h_{f}, J_{3f}) t]$. By taking into account the overlap between evolved state and the initial state, we calculate the complexity using Eq. (\ref{s1quench}).
 \item  In the next step, the initial state will be evolved by $U_{2}(t)=\exp[-i H_{i}(h_{i},J_{3i})t]\exp[-i H_{f}(h_{f},J_{3f})t_{1}]$, where $t_{1}=10$ and $10<t\leq 20$ and again by taking overlap of evolved state with an initial state, complexity can be calculated.
\item In the last step, the initial state evolved by $U_{3}(t)=\exp[-i H_{f} (h_{f},J_{3f}) t]\exp[-i H_{i} (h_{i},J_{3i}) t_{2}]\exp[-i H_{f} (h_{f},J_{3f}) t_{1}]$, where $t_{1}=10$, $t_{2}=20$ and $20 <t\leq 50 $,
and the overlap of the initial state with the evolved state provide us complexity.
\end{itemize}
The results for the time evolution of the SC are depicted in Fig. \ref{fig:multi}. 
We consider behaviour around one of the critical points. The red curve shows  the time evolution of SC
when the final parameters are at a critical point and the blue curve when initial parameters are at critical points. In Fig. \ref{fig:multi}, we take $h=J_{3}-1$ as the critical point, and for the 
blue curve, the parameter values are $h_{i}=0.6$, $J_{3i}=1.6$, $h_{f}=1$, $J_{3f}=1.2$,  while for the red curve, we take $h_{i}=1$, $J_{3i}=1.2$, $h_{f}=0.6$, $J_{3f}=1.6$.
Analysing these results, we can see the amplitude of complexity depends on the initial and final values of the parameters and that, with increasing time, it attains steady values. 
Although the amplitude decreases by switching the value of parameters, there is no specific behaviour at the QPT. 
The behaviour of SC for multiple quenches is almost similar to the case of a single quench.

 \subsection{Spread complexity evolution with periodically driven field}
In this section, we will study how the SC behaves when the system is driven by periodically varying fields \cite{russo,vic,shrad} over 
stroboscopic time ($t=n T$, where $n\in \mathbb{Z}$). We replace $h=h(t)=\delta \cos{\omega t}+h_{c}$ in Eq. (\ref{three hamilt}), 
$T$ is defined as the period for one oscillation $T=\frac{2\pi}{\omega}$. In this case, we use the invariant operator technique \cite{lai} or the 
time-dependent unitary transformations \cite{Maam} $i\frac{d}{dt}\ket{\psi(t)}=H(t)\ket{\psi(t)}$, $\hat{J_{0}}\ket{n}=J_{n}\ket{n}$. According 
to the well-known Lewis-Riesenfeld theory \cite{Lewis}, the general solution of the Schrodinger equation is written as, $\ket{\psi(t)}=\sum_{n}c_{n}\exp(i\alpha_{n}(t))\ket{n,t}$, where  $\ket{n,t}$ are eigenstate of the invariant operator
$\mathcal{I}(t)\ket{n,t}=J_{n}\ket{n,t}$, with $\ket{n,t}=\hat{R}(t)\ket{n}$.
This can also be written in the form 
 \begin{equation}
    \ket{\psi (t)}=\hat{R}(t)\exp[-i\epsilon(t)J_{0}]\hat{R}^{\dagger}(0)\ket{\psi(0)}
    \label{tpstate}~,
\end{equation}
where we have denoted 
\begin{equation}
\hat{R}(t)=\exp\Big[\frac{\gamma(t)}{2}\Big(\hat{J}_{+}e^{-i\beta(t)}-\hat{J}_{-}e^{i\beta(t)}\Big)\Big]~,
 ~\text{and}~~
 \gamma(t)=\tan^{-1}\Big[\frac{|(J_3 \sin{2k}-\sin{k})|}{\cos{\omega t}+h_{c}+\cos{k}-J_3 \cos{2k}}\Big]~.\label{rt1}
\end{equation}
 
We have to solve an auxiliary equation to find out the function $\gamma(t)$ and $\beta (t)$. Here we outline 
the computation of the return amplitude, as well as  SC, while some of the relevant mathematical
details are provided in Appendix \ref{timede1} and Appendix \ref{timede2} (see also \cite{lai}).
To calculate SC we
will express Eq. (\ref{tpstate}) in terms of the Krylov basis, which can be written by using the BCH 
 formula \cite{Bch}. And to calculate the SC, we take the initial state to be the ground state, i.e., 
\begin{equation}\label{3initial}
\ket{\psi(0)}=\hat{R_{i}}\ket{j,-j}~,~~\hat{R_{i}}=\exp\Big[-i\frac{\pi-\phi_{i}}{2}(J_{+}+J_{-})\Big]
\end{equation}
where the explicit expression for $\phi_{i}$ is given by
 \begin{equation}
  \phi_{i}=\cos^{-1}\Big[\frac{(h_{c}+\cos{k}-J_3\cos{2k})}{\sqrt{(h_{c}+\cos{k}-J_3\cos{2k})^{2}+(J_3\sin{2k}-\sin{k})^{2}}}
  \Big] ~.   
 \end{equation}
The expression for the  return amplitude for a single mode of momentum mode for general $j$ is given by 
 \begin{equation}\label{return1/2}
     \mathcal{S}=\Big[\cos\Big[\frac{\gamma(0)-\gamma(t)}{2}\Big]\cos\big[\frac{\epsilon(t)}{2}\big]
     -i\cos\Big[\frac{\gamma(0)+\gamma(t)}{2}-\phi_{i}\Big]\sin\big[\frac{\epsilon(t)}{2}\big]\Big]^{2j}~,
 \end{equation}
where
\[
\epsilon(t)=\int_{0}^{t}\sqrt{(h(t')+\cos{k}-J_3\cos{2k})^{2}+(J_3\sin{2k}-\sin{k})^{2}}dt'~.
\]
$\gamma(t)$ and $\hat{R}(t)$ are periodic, such that $\gamma (0) = \gamma (nT)$ and $\hat{R} (0) = \hat{R} (nT)$ \cite{Maam}.
Therefore, for periodic condition $\gamma(nT)=\gamma(0)=\gamma$, the SC for a single mode in free fermion $j-\frac{1}{2} $ case given by 
 \begin{equation}
     \mathcal{C}_k= 1- |\mathcal{S}|^{2}=
     \sin\Big[\frac{\epsilon(nT)}{2}\Big]^{2}\sin[\gamma-\phi_{i}]^{2}~.
\end{equation}
 As before,  in the continuum limit, we obtain the total complexity to be 
 \begin{equation}
           \mathcal{C}(s=1)= \frac{1}{2\pi} \int_{0}^{\pi}(1-|\mathcal{S}|^{2})dk \label{periodspr}~.
\end{equation}

\begin{figure}[h!]
	\centering
	\subfigure[]{
		\includegraphics[width=0.3\textwidth]{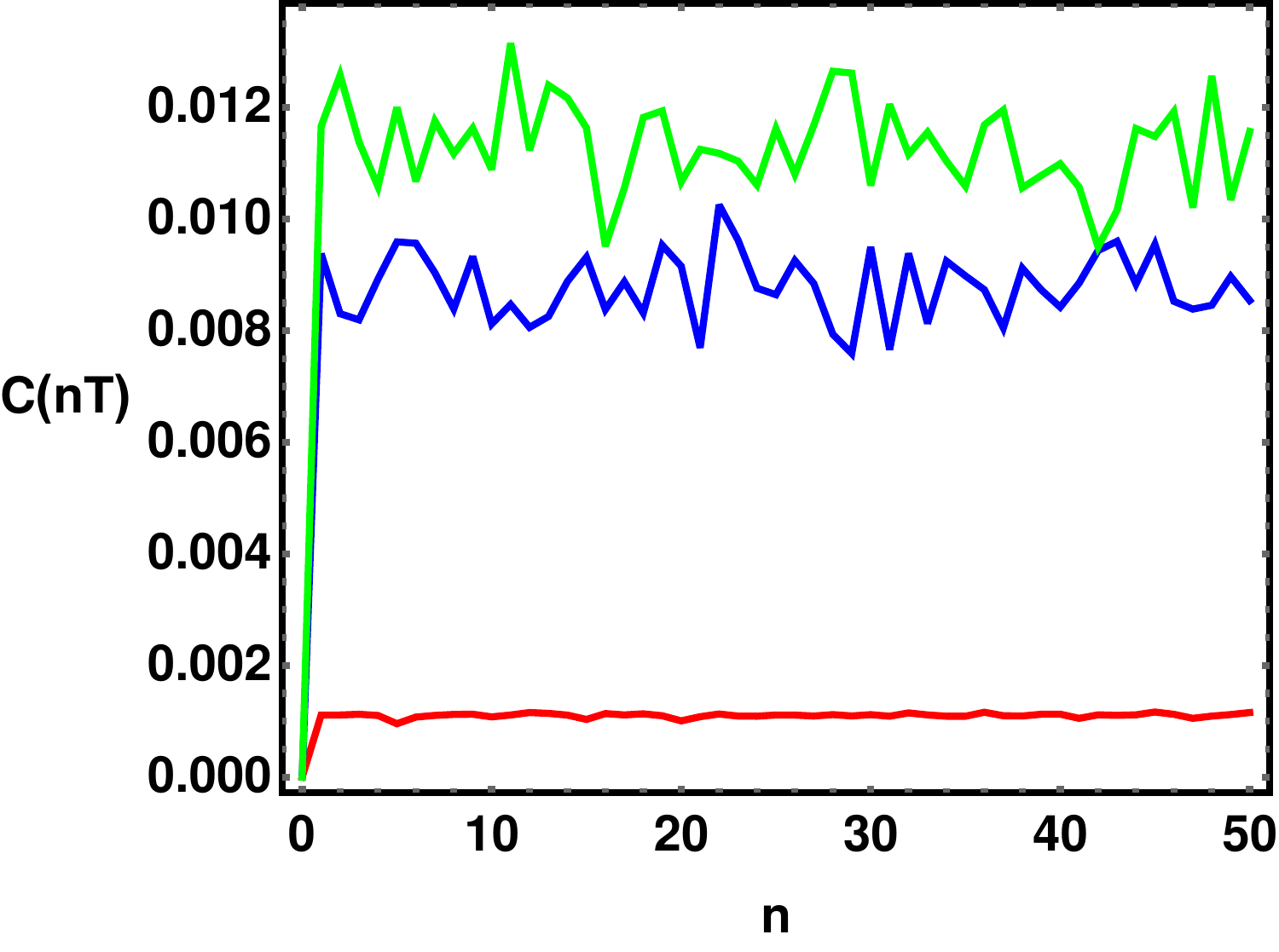}
		\label{fig:osc3}}
	\centering
	\subfigure[]{
		\includegraphics[width=0.3\textwidth]{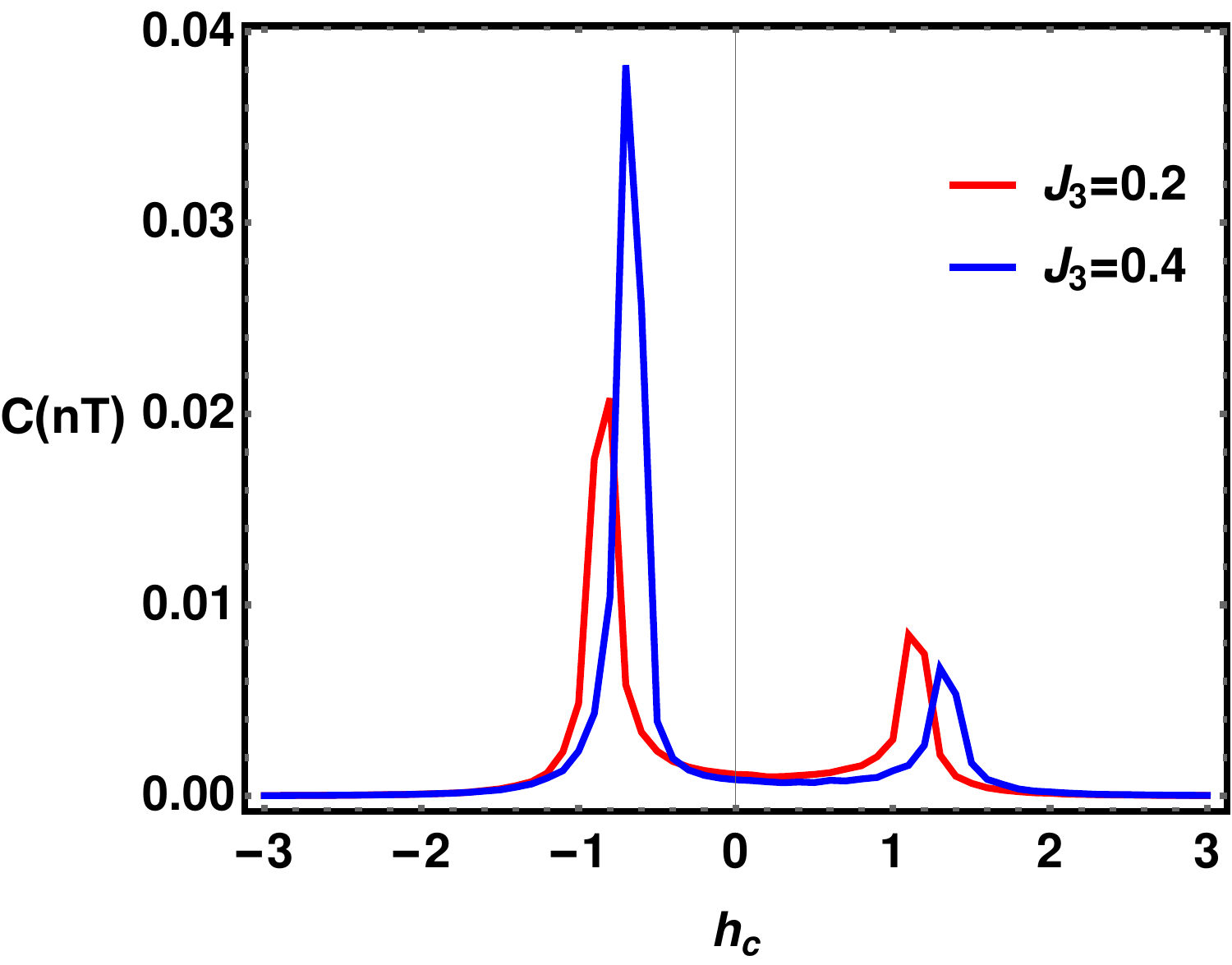}
		\label{fig:3qpta}}
	\centering
	\subfigure[]{
		\includegraphics[width=0.3\textwidth]{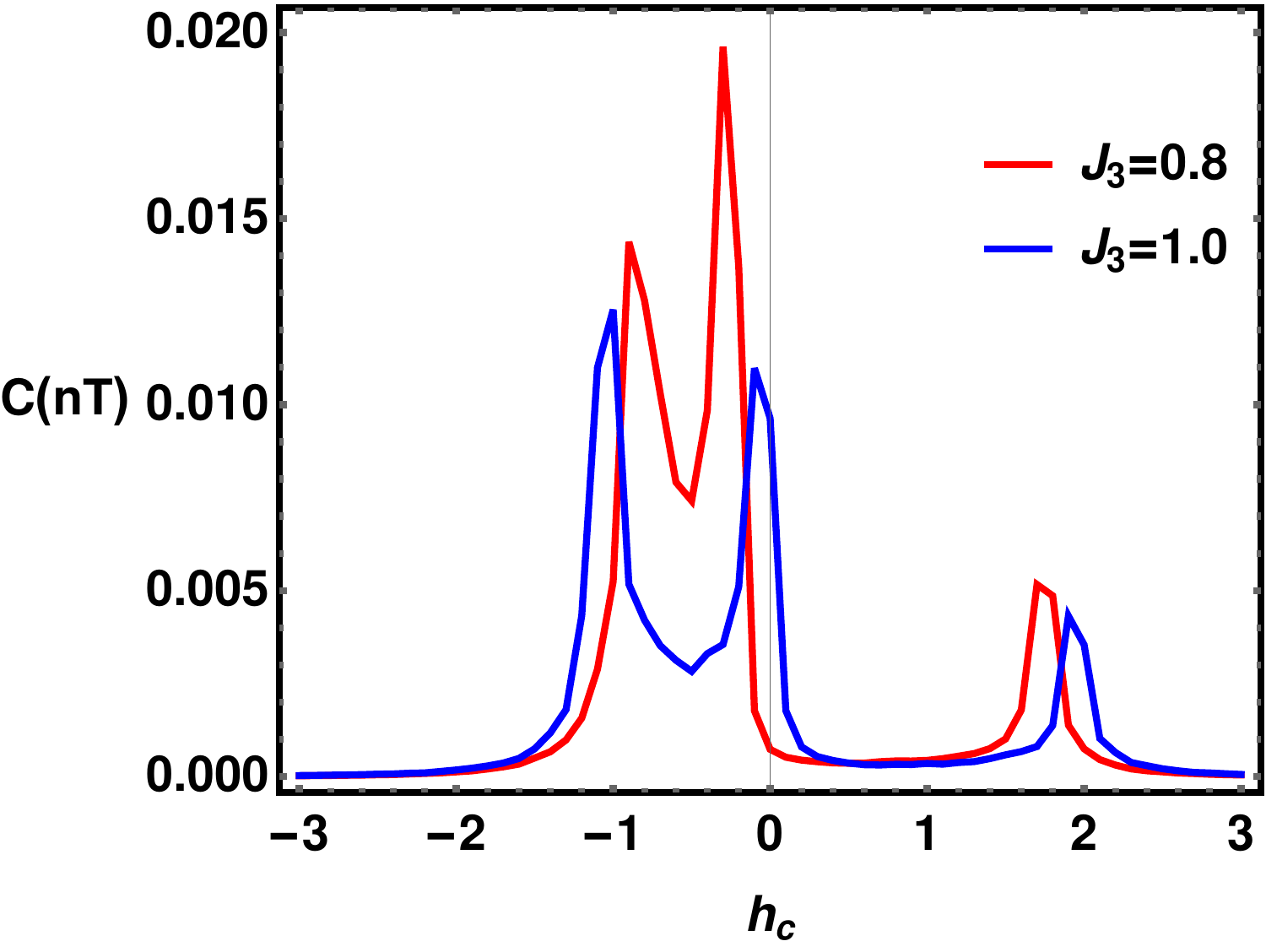}
		\label{fig:3qptb}}
	\caption{SC evolution  in the case of periodically driven magnetic field (a) Complexity variation with number of oscillation, (b) Showing the 
		two peaks that correspond to $h=J_{3}+1 $ and $h=J_{3}-1$, (c) Showing the three peaks at $h=J_{3}+1$, $h=J_{3}-1$, $h=-J_3$.}
\end{figure}

We calculate complexity in Eq. 
(\ref{periodspr}) by using numerical integration method of Simpson- $\frac{1}{3}$ over $k$ from  $(0,\pi)$  with step size $\frac{\pi}{1000}$, while doing the integration we fix $\delta=0.1$. We plot Eq. (\ref{periodspr}) with respect to number of oscillation $t=nT$, $T=\frac{2\pi}{\omega}$, $n\in \mathbb{Z}$. In Fig. \ref{fig:osc3} we have fixed $T=1000$ and vary the parameters $J_3$, $h_c$, as red: $J_3=0.2$, $h_c=0$, blue: $J_3=0.2$ , $h_c=1.1$, and green: $J_3=1$, $h_c=-1.1$. For the first cycle in Fig. \ref{fig:osc3}, complexity growth is at its maximum; from then on, it begins to oscillate around an average value.
 We plot Eq. (\ref{periodspr}) with respect to $h_{c}$ by taking different values of $J_3$ and keeping $n=40$, $T=1000$. In Fig. \ref{fig:3qpta} we take $J_3<0.5$,  two peak correspond $h(nT)=J_3+1$ and $h(nT)=J_3-1$. In Fig. \ref{fig:3qptb} we take $J_3>0.5$ and three peaks correspond to $h(nT)=J_3+1$, $h(nT)=J_3-1$, $h(nT)=-J_3$.

\section{XY Spin Model}\label{xyspin}
In this section, we consider the spin-$\frac{1}{2}$ XY model  placed in a magnetic field ($h$) pointed along $\hat{z}$ direction \cite{sharma}.
The Hamiltonian, in this case, is given by 
\begin{equation}
    H=-\sum_{j=1}^{N-1}\Big[\frac{1+\gamma}{2}\sigma_{j}^{x}\sigma_{j+1}^{x}+\frac{1-\gamma}{2}\sigma_{j}^{y}\sigma_{j+1}^{y}\Big]-\sum_{j=1}^{N}h\sigma_{j}^{z}~.
    \label{hamit xy}
\end{equation}
where $\gamma$ is the anisotropy parameter. This model and Three spin interaction TIM Eq. (\ref{three hamilt}) are related by a duality transformation (see Appendix \ref{dutr}). 
As before by performing successive  Jordan-Wigner  \cite{Jw1,Jw2,Jw3} and Fourier transformations we rewrite the above 
Hamiltonian as
\begin{equation}
    H=-\sum_{k>0}(h+\cos{k})(c^{\dagger}_{k}c_{k}-c_{-k}c^{\dagger}_{-k})+\sum_{k>0}
    \big[i\gamma c^{\dagger}_{k}c^{\dagger}_{-k}-i\gamma c_{-k}c_{k}\big]~.
    \label{xyjw}
\end{equation}
where now the energy gap is given by
\begin{equation}
    \epsilon={\sqrt{(h+\cos k)^{2}+(\gamma \sin k)^{2}}}~.
    \label{two gap}
\end{equation}
Eq. (\ref{two gap}) has extrema at $k=0$, $k=\pi$ and $\cos k_{o}=\frac{h}{-1+\gamma^{2}}$. 
Using these values of $k$ we can find out that Eq. (\ref{two gap}) vanishes at $h=\pm 1$ and $\gamma=0$ (provided $|h|<1$). Therefore critical points for this system are $h=\pm 1$ and $\gamma=0$ for $h<1$.

Hamiltonian Eq. (\ref{xyjw})  can be written  as an element of the $su(2)$ algebra as (see Appendix \ref{xyapp} for derivation)
\begin{equation}
   H=\sum_{k>0}\Big[-2 R \cos\phi J^{k}_{0}+i R \sin\phi J^{k}_{+}-i R \sin\phi J^{k}_{-}\Big]~,
   \label{xysu2R}
\end{equation}
where we have defined 
\begin{equation}\label{xydef}
R(k)=\sqrt{R_2^{2}+R_3^{2}}~, \qquad R_2=\gamma\sin{k},\qquad R_3=h+\cos{k} ~, \qquad \phi=\tan^{-1}\frac{|R_2|}{R_3}   ~.
\end{equation}

\begin{figure}[h!]
	\centering
	\subfigure[]{
		\includegraphics[width=0.3\textwidth]{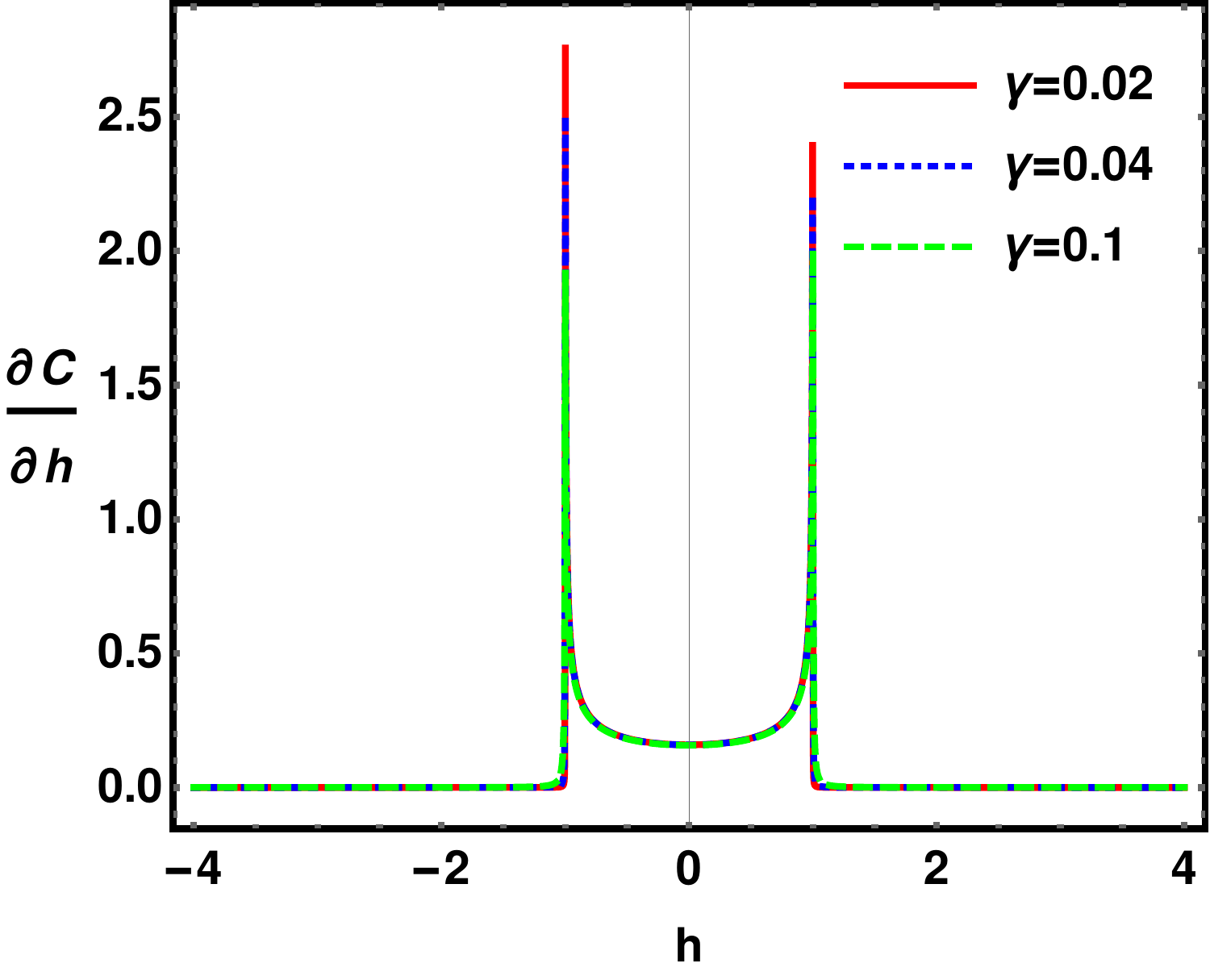}
		\label{fig:drvxy}}
	\centering
	\subfigure[]{
		\includegraphics[width=0.3\textwidth]{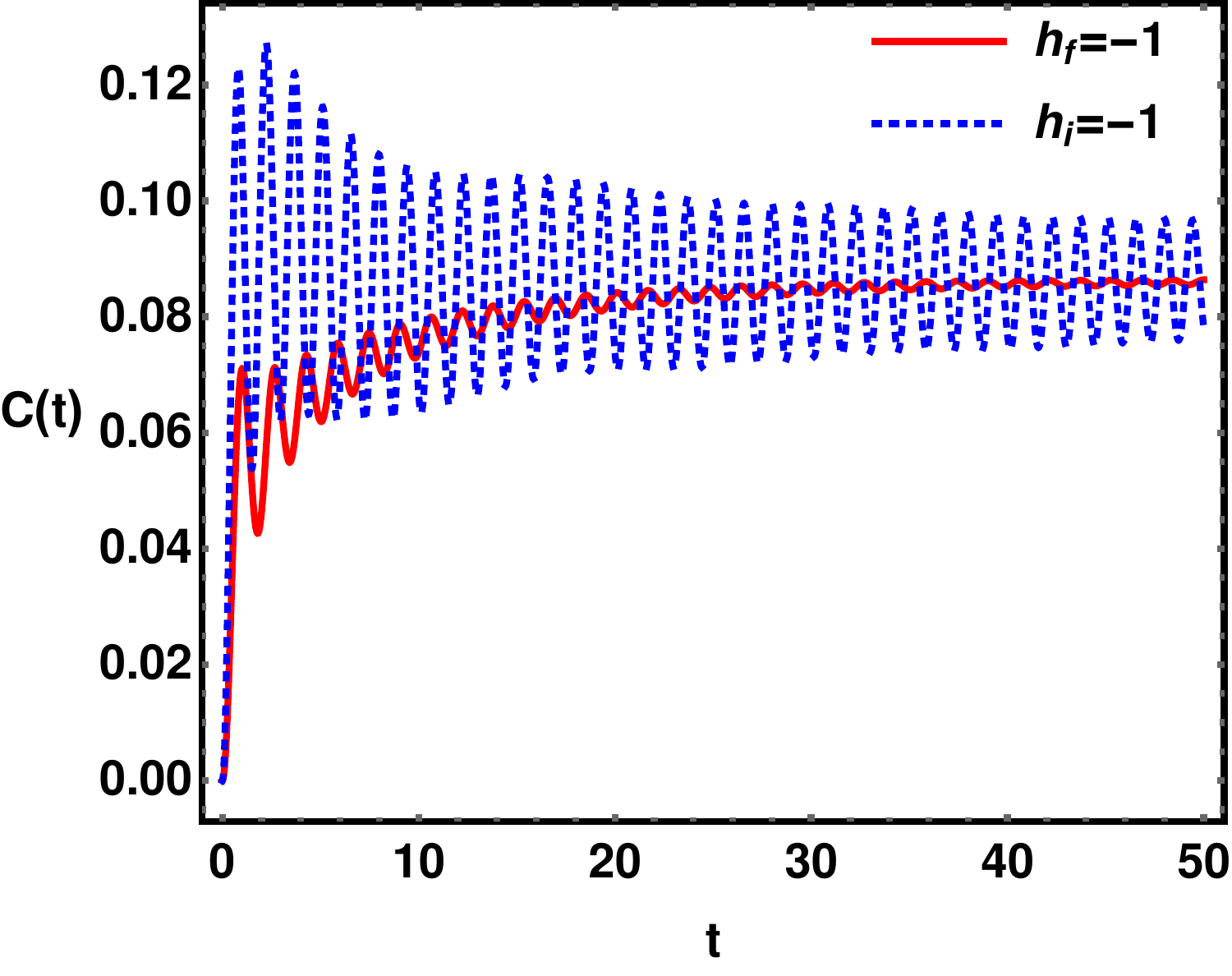}
		\label{fig:squench2xy}}
	\centering
	\subfigure[]{
		\includegraphics[width=0.3\textwidth]{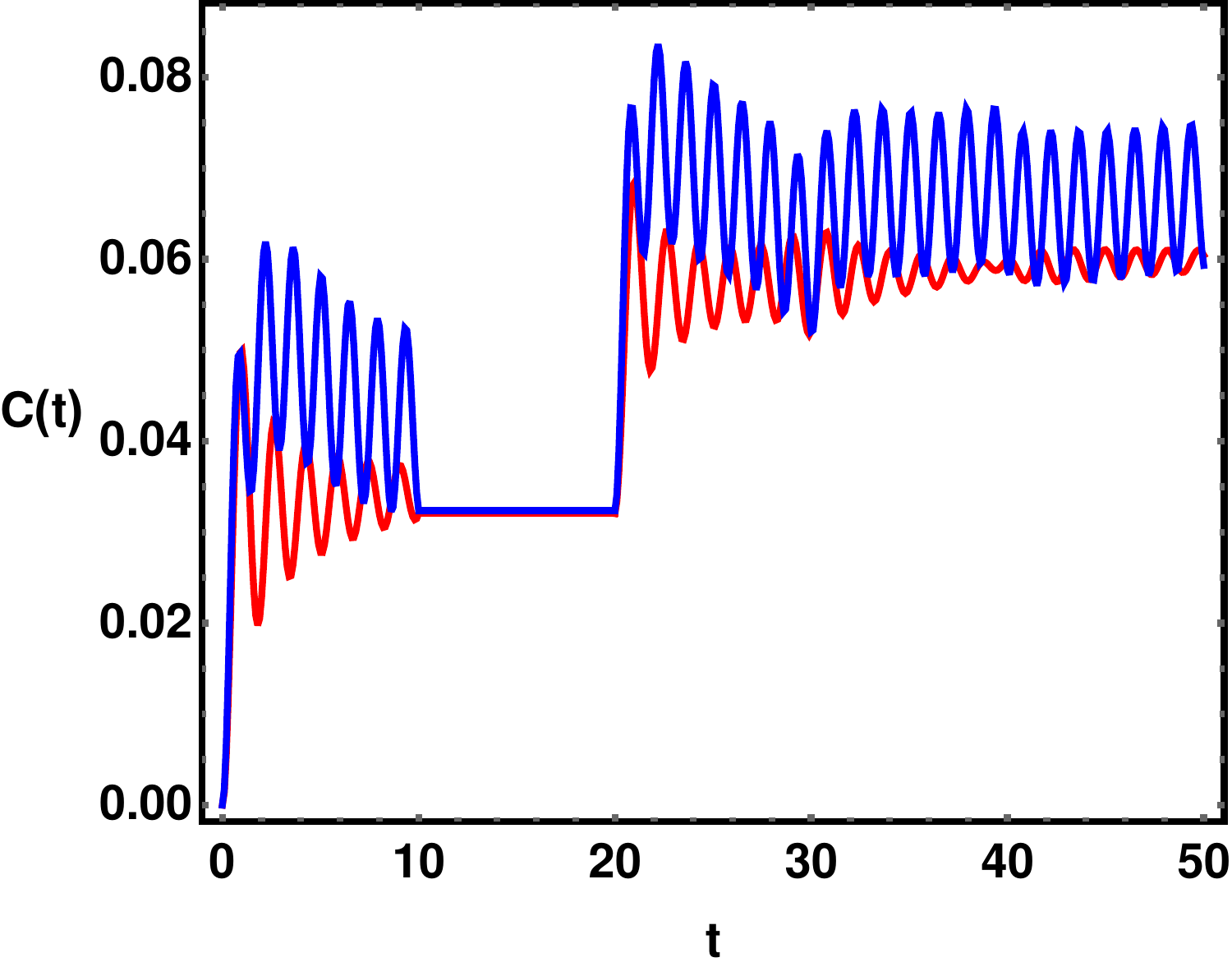}
		\label{fig:mquench2xy}}
	\caption{\small{Spread complexity of the XY spin chain model : (a)\, Derivative of complexity with respect to $h$. (b)\, Complexity for a single quench. (c)\, Complexity for multiple quenches}}
	\label{fig:}
\end{figure}

The ground state of this model is of  the same form as the three-spin interaction Ising model considered in the 
the previous section, i.e. 
\begin{equation}
\ket{\psi_{gs}}=\prod_{k>0}\cos{\frac{\pi-\phi}{2}}\exp\Big[-i\tan{\frac{\pi-\phi}{2}}J_{+}^{k}\Big]\ket{\frac{1}{2},-\frac{1}{2}}_{k}~,
\end{equation}
This  can be rewritten as
\begin{equation}
\ket{\psi_{gs}}=\prod_{k>0}\exp\Big[-i\frac{\pi-\phi}{2}J_{+}^{k}-i\frac{\pi-\phi}{2}J_{-}^{k}\Big]
\ket{\frac{1}{2},-\frac{1}{2}}_{k}~,
\end{equation}
SC of formation of the ground state is given by the same formula in Eq. (\ref{com three}), with 
$\phi$, in this case, is the one given in  Eq. \eqref{xydef}.
In Fig. \ref{fig:drvxy}, we have plotted the derivative of the SC in this case. As can be seen the
derivative of SC  shows non-analytical behaviour around $h=-1$ and $h=1$.

\subsection{Spread Complexity evolution after a single quench}

The quench protocol we consider here is the same as the one considered  in the three-spin interactions Ising case. We consider a quench
in this model such that the
initial parameter values $h_{i}$, $\gamma_{i}$, are changed to new set of values $h_{f}$, $\gamma_{f}$. 
The SC can be calculated using the same procedure employed in the previous section and in the continuum limit
the expression for the complexity is the same as in  Eq. \eqref{s1quench}, i.e. 
\begin{equation}
\mathcal{C}(t)=\frac{1}{2\pi}\int_{0}^{\pi}\sin^{2}(\phi_{f}-\phi_{i})\sin^{2}(R_{f}t)dk ~,
\end{equation}
where the expression for $\phi_{f}$ and $\phi_{i}$ are determined by the formula 
for $\phi$ in Eq. \eqref{xydef}.

In Fig. \ref{fig:squench2xy}, we have plotted the complexity evolution after a single quench. 
We consider the case when parameters are at one of the critical points $h=-1$,  where the red curve reflects the behaviour of final parameters at the critical point, and the blue curve represents when initial parameters are at the critical point.
For red curve we choose the parameter values : $h_{i}=1.2$, $\gamma_{i}=0.4$, $h_{f}=-1.0$, $\gamma_{f}=0.2$, blue: $h_{i}=-1.0$, $\gamma_{i}=0.2$, $h_{f}=1.2$, $\gamma_{f}=0.4$.

\subsection{Spread complexity evolution after multiple quenches}
In multiple quenches, the quench protocol is the same as in three spin cases; here again, we consider one of the critical points $h=-1$. In Fig., \ref{fig:mquench2xy} red curve reflects the behaviour for the final parameter at a critical point and blue for the initial parameter at a critical point.
The parameter values are for the blue curve are : $h_{i}=-1$, $\gamma_{i}=0.05$, $h_{f}=1.2$, $\gamma_{f}=0.4$ 
and for red curve parameters are $h_{i}=1.2$, $\gamma_{i}=0.4$, $h_{f}=-1$, $\gamma_{f}=0.05$. For the case
when the final parameters are at the critical point, complexity has fewer oscillations, and the magnitude of amplitude is less as compared to the case when the initial parameter is  at a critical point. This can be understood from the expression of SC where we have time-dependent factor $(\sin(R_{f} t))^{2}$. Now since $R_{f}$ has the form of Energy gap $R_{f}=\sqrt{(h_{f}+\cos k)^{2}+(\gamma_{f}\sin k)^{2}}$, this attains minimum value at the critical points.

\subsection{Spread complexity evolution with periodically driven field}
In this case, we consider  $h(t)=\delta \cos{(\omega t)}+h_c$  where time $t$ is discrete and measured in terms of integer $n$, $t=\frac{2n\pi}{\omega}$. 
The expression for the time evolved state  is the same as Eq. \eqref{tpstate} 
where now we have  \footnote{Note that $\gamma(t)$ is different from the $\gamma$ we used to represent the anisotropy parameter
	in this model.} 
\begin{equation}
\hat{R}(t)=\exp\Big[\frac{i\gamma (t)}{2}(J_{+}^{k}+J_{-}^{k})\Big]~,~ ~~\text{and}~~
\gamma(t)=\tan^{-1}\Big[\frac{\gamma \sin k}{h(t)+\cos k}\Big]~.
\end{equation}
We calculate the evolution of the SC of this time evolved state with respect to an initial state 
$\ket{\psi(0)}=\hat{R_{i}}\ket{j,-j}$ (same as Eq. \eqref{3initial}) 
with 
 \[\phi_{i}=\cos^{-1}\Big(\frac{(h_{c}+\cos{k})}{\sqrt{(h_{c}+\cos{k})^{2}+(\gamma\sin{k})^{2}}}\Big)~.\]

With periodic boundary conditions, the expressions for the return amplitude  for a single momentum mode and the complexity are the same as the ones in Eqs.
\eqref{return1/2} and \eqref{periodspr} respectively, where now the expression for the time-dependent 
function $\epsilon(t)$ is given by
\begin{equation}
\epsilon(t)=\int_{0}^{t}\sqrt{(h(t')+\cos{k})^{2}+(\gamma\sin{k})^{2}}dt'~.
\end{equation}

\begin{figure}[h!]
    \centering
    \subfigure[]{
 \includegraphics[width=0.3\textwidth]{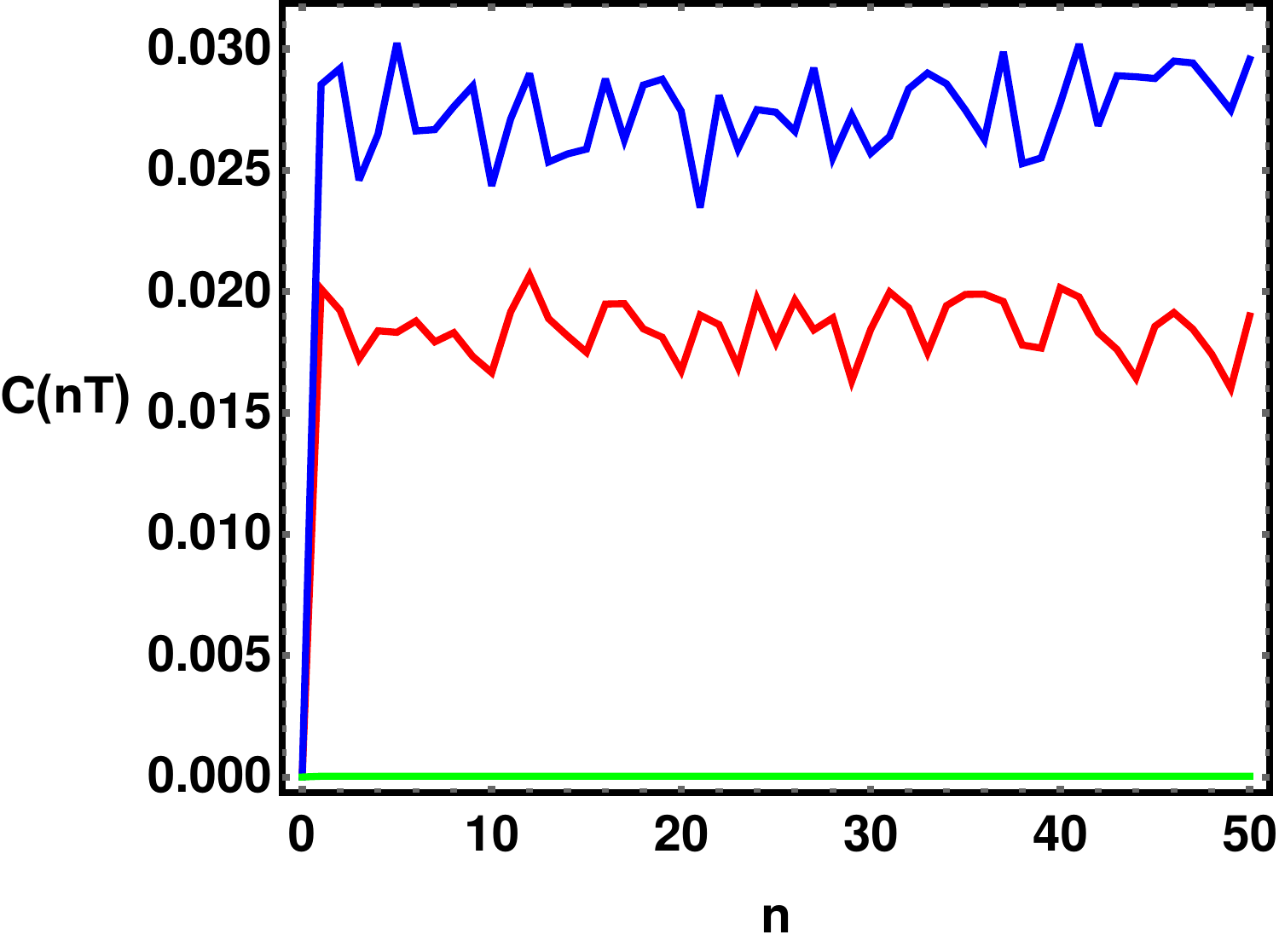}
    \label{fig:periodicspra}}
    \centering
    \subfigure[]{
 \includegraphics[width=0.3\textwidth]{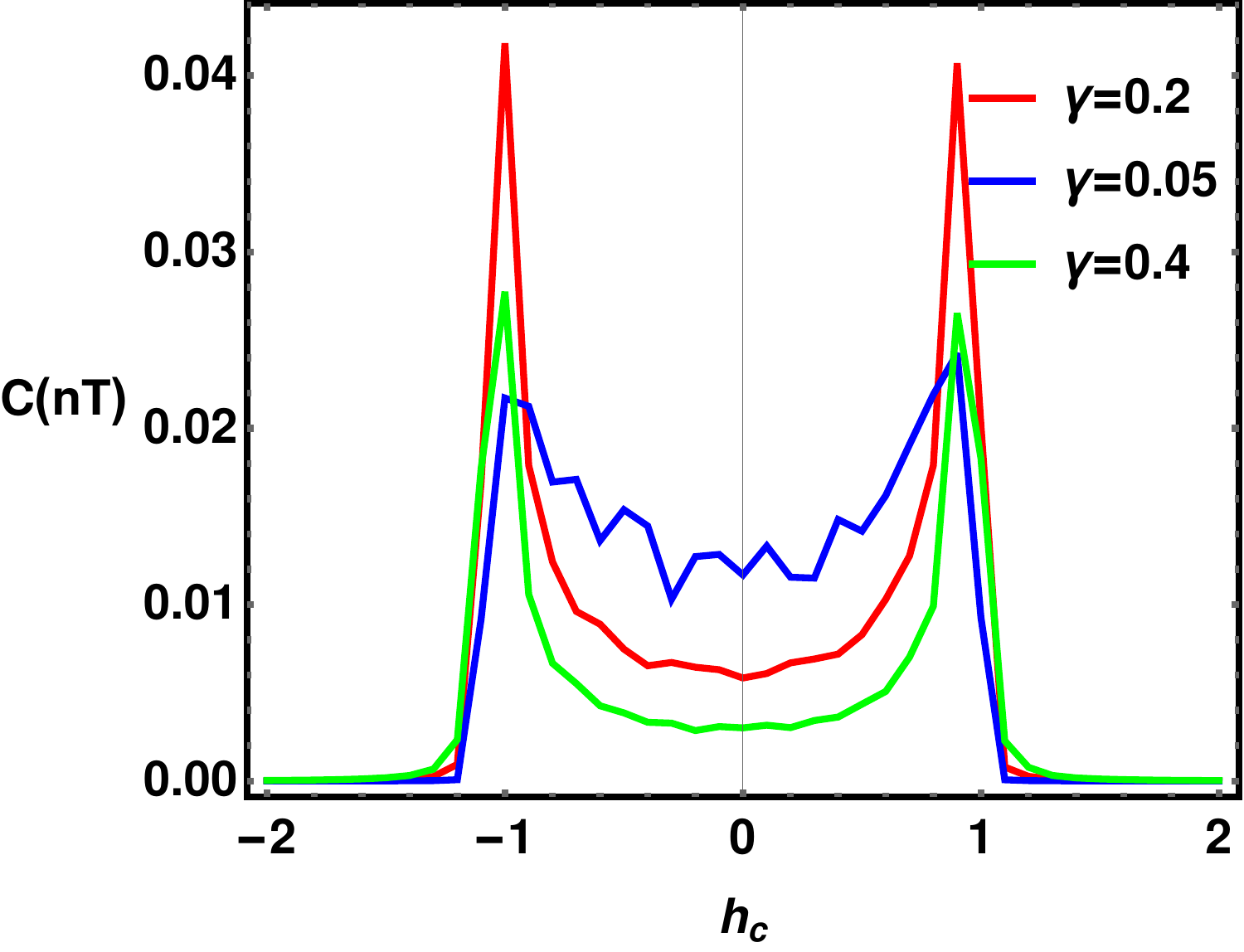}
    \label{fig:periodicsprb}}
    \caption{\small{SC evolution with the periodic field in XY spin chain model : (a)\, Variation  with a number of oscillations. (b)\, Variation  with parameter $h_c$. }}
    \label{fig:periodicspr}
\end{figure}

 Fig. \ref{fig:periodicspr}  shows  the time evolution of the SC with $\delta=0.1$. In Fig. \ref{fig:periodicspra},
 we have plotted it with  a number of oscillations $n$ with following  values of the parameters: red  $:\gamma=0.2$, $h_c=1$,
  blue $:$ $\gamma=0.4$,  $h_c=-1$, and  green: $\gamma=0.2$,  $h_c=1.5$.
On the other hand, in Fig. \ref{fig:periodicsprb}, we have plotted it with respect to $h_c$ by taking $n=40$, $T=1000$ in all 
 three cases.

\section{Su-Schrieffer-Heeger Model}
The SC in the SSH model  has been studied previously in \cite{captua} in the context
of a single quantum quench.  Here we are interested in studying the SC in
this model after multiple quenches and in the presence of a time-varying
field. Physically speaking, it defines a 1D lattice with a two-atom sublattice structure in which the hopping amplitudes between the lattices are typically different. 
\\

Hamiltonian of the SSH model is given by :
\begin{equation}
H=t_1\sum_{i}(c_{Ai}^{\dagger}c_{Bi}+H.c)-t_2\sum_{i}(c_{Bi}^{\dagger}c_{A,i+1}+H.c)+\mu_{s}\sum_{i}(c_{Ai}^{\dagger}c_{Ai}-c_{Bi}^{\dagger}c_{Bi})\label{sshhamilt}
\end{equation}
where $t_1$ is intra-cell hopping amplitude and $t_2$ is inter-cell hopping amplitude. $c_{Ai}$, $c_{Bi}$ represent fermions annihilation operators defined at site $i$ for sub lattice $A$ and $B$ respectively. We take $t_1, t_2\geq0$ and $\mu_{s}=0$. Depending on the values of parameters, there are two phases: non-topological phase for $t_1>t_2$ and topological phase for $t_2>t_1$.
This model shows a QPT at $t_1=t_2$. 
Eq. (\ref{sshhamilt}) can be diagonalized using Fourier transform and Bogoliubov transformations.

In momentum space (with $\mu_{s}=0)$ the Hamiltonian  can be rewritten as 
\begin{equation}
H=\sum_{k}\Big[2 R_3 J_{0}^{k}+i R_{1}(J_{+}^{k}-J_{-}^{k})\Big]~,
\label{sshsu2}
\end{equation}
where 
\begin{equation}\label{sshdef}
R_{1}=t_{1}-t_{2}\cos k~, \quad  R_{3}=t_{2}\cos k~, \quad ~~\text{and}~ R(k)=|\vec{R}(k)|~.
\end{equation}
 The operators $J_{0}^{k},J_{\pm}^{k}$
satisfy the usual $su(2)$ Lie algebra, and future convenience we define 
\begin{equation}
\sin\phi=\frac{|R_{1}|}{R}~,\quad \cos{\phi}=\frac{R_3}{R}~, \text{with}~~R=\sqrt{t_{1}^{2}+t_{2}^{2}-2 t_1 t_2 \cos k}~.
\label{parar1r2ssh}
\end{equation}

The ground state of this model is given by
\begin{equation}
    \ket{\psi_{g}}=\prod_{k>0}\cos^{2}\Big(\frac{\phi}{2}\Big) \exp\Big[-i\tan\Big(\frac{\phi}{2}\Big)\big[J_{+}^{k} e^{-i\psi_{k}}+J_{-}^{k}e^{i\psi_{k}}\big]\Big]\ket{\frac{1}{2},\frac{-1}{2}}_k~.
    \label{grndst}
\end{equation}
The SC of the ground state  in the continuum limit is given by  \cite{captua}
\begin{equation}
\mathcal{C}=\frac{1}{\pi}\int_{0}^{\pi}dk \left(\sin{\frac{\phi}{2}}\right)^{2} ~.
\end{equation}

\begin{figure}[h!]
 \centering
    \subfigure[]{
 \includegraphics[width=0.3\textwidth]{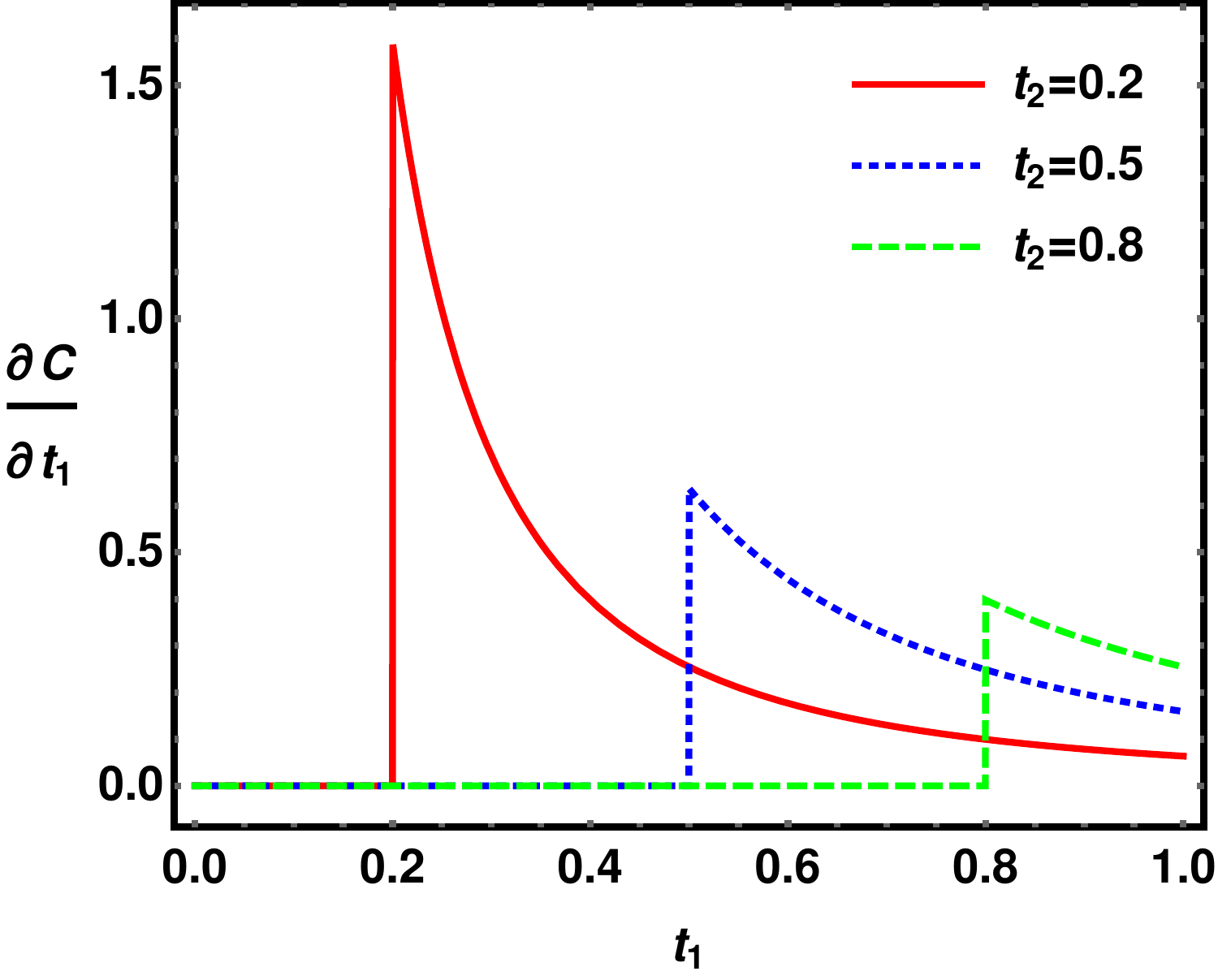}
\label{dervssh}}
\centering
    \subfigure[]{
 \includegraphics[width=0.3\textwidth]{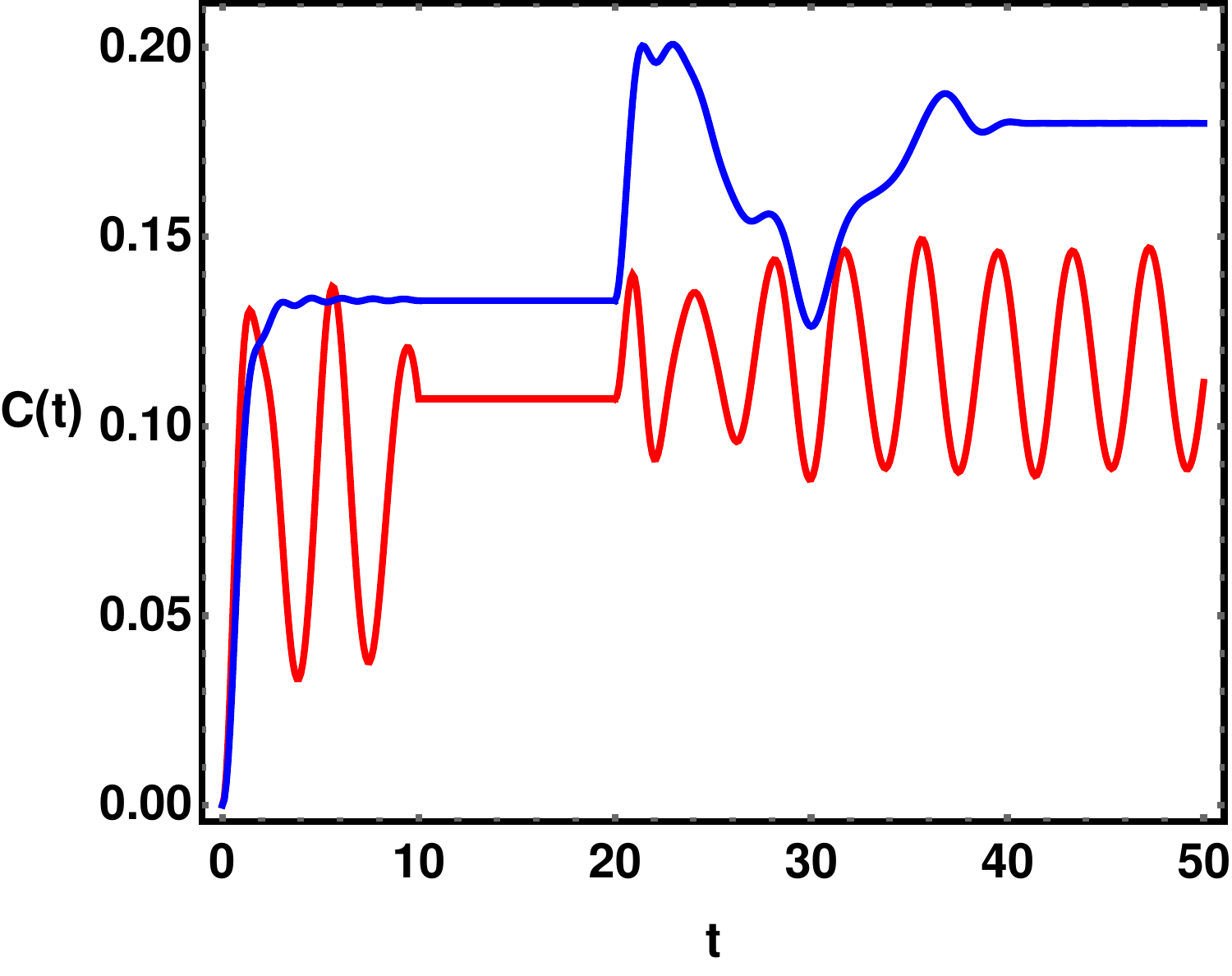}
    \label{multissh}}
    \caption{\small{SC in  SSH model : (a)\, Derivative of complexity with parameter $t_{1}$ (b)\, Evolution for 
    		multiple quenches.}}
    \label{spreadssh}
    \end{figure}

\subsection{Complexity evolution after multiple quenches}
The time evolved state after a quench is given by
\begin{equation}
    \ket{\psi(t)}=e^{-iH_{f} t}\ket{\psi_{i}}~.
\end{equation}
where $\ket{\psi_{i}}$ is the ground state defined in Eq. (\ref{grndst}). Note  that here  $\phi=\phi_{i}$ and $\psi_{k}=0$, since $\mu_s=0$. In Fig. \ref{dervssh}, we plot the derivative of spread complexity with respect to the parameter without quench case, and it clearly shows a non-analytical nature near the critical point $t_{1}=t_{2}$.

Return amplitude after a single quench can be written as  \cite{captua}
\begin{equation}
\mathcal{S}=\braket{\psi(t)|\psi(0)}=\cos(R_{f} t)-i\cos(\phi_{f}-\phi_{i})\sin(R_{f}t)~.
\end{equation}
The SC calculated from this return amplitude is given by 
\begin{equation}
 \mathcal{C}(t)=\int_{0}^{\pi}\frac{dk}{\pi}\sin^{2}(\phi_{f}-\phi_{i})\sin^{2}(R_f t)~.
 \label{quenssh}
\end{equation}
In Fig. \ref{multissh}, we have plotted the behaviour of SC under multiple quenches,
where the parameter values for the red curve: $t_{1i}=1=t_{2i}$,  $t_{1f}=0.7$,  $t_{2f}=1.5$, and for the blue curve: $t_{1i}=1.5$, $t_{2i}=0.7$, $t_{1f}=1=t_{2f}$, where $i$ subscript represents initial value and $f$ represents final value.

\begin{figure}[h!]
	\centering
	\subfigure[]{
		\includegraphics[width=0.3\textwidth]{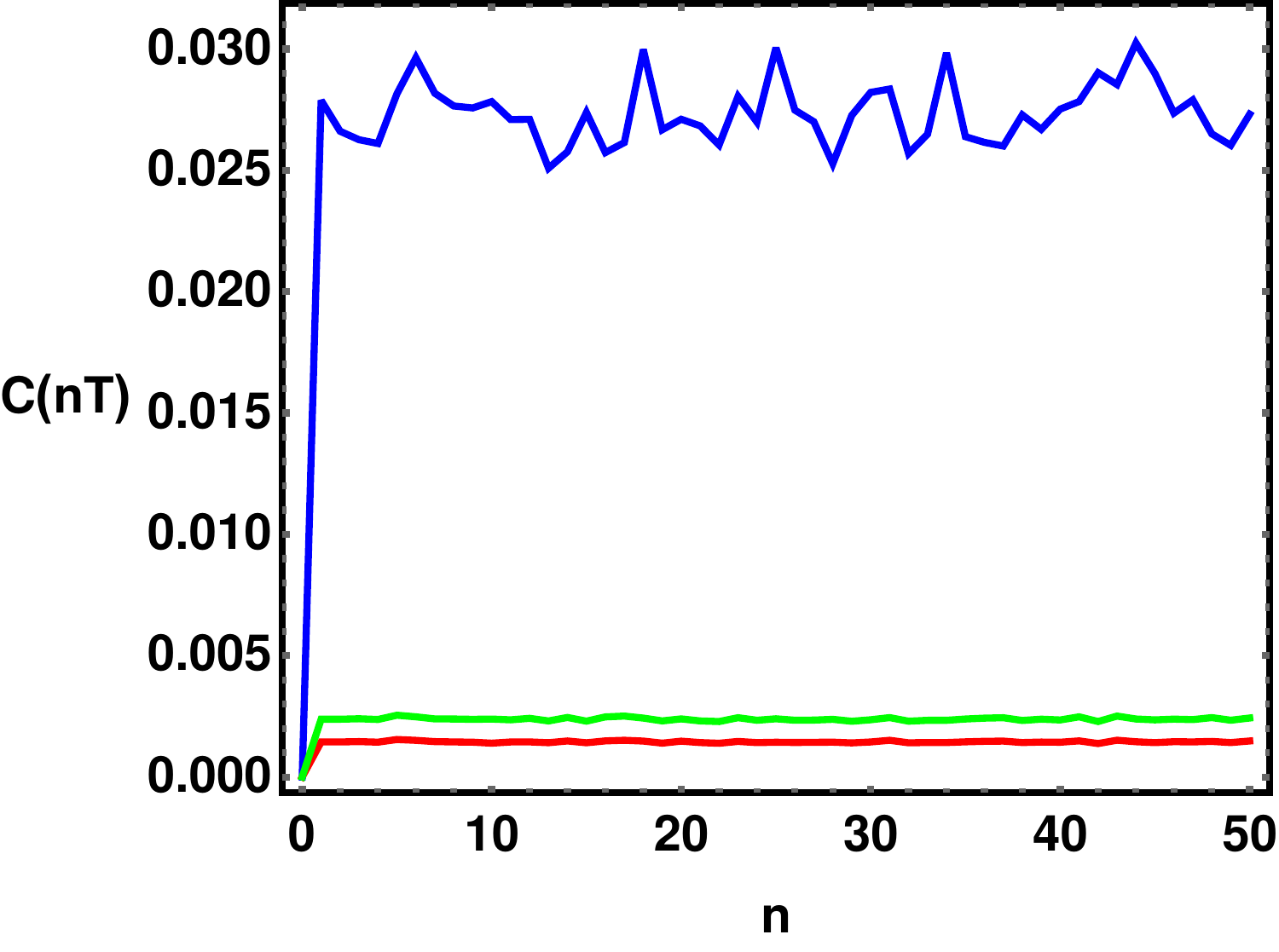}
		\label{sshoscn}}
	\centering
	\subfigure[]{
		\includegraphics[width=0.3\textwidth]{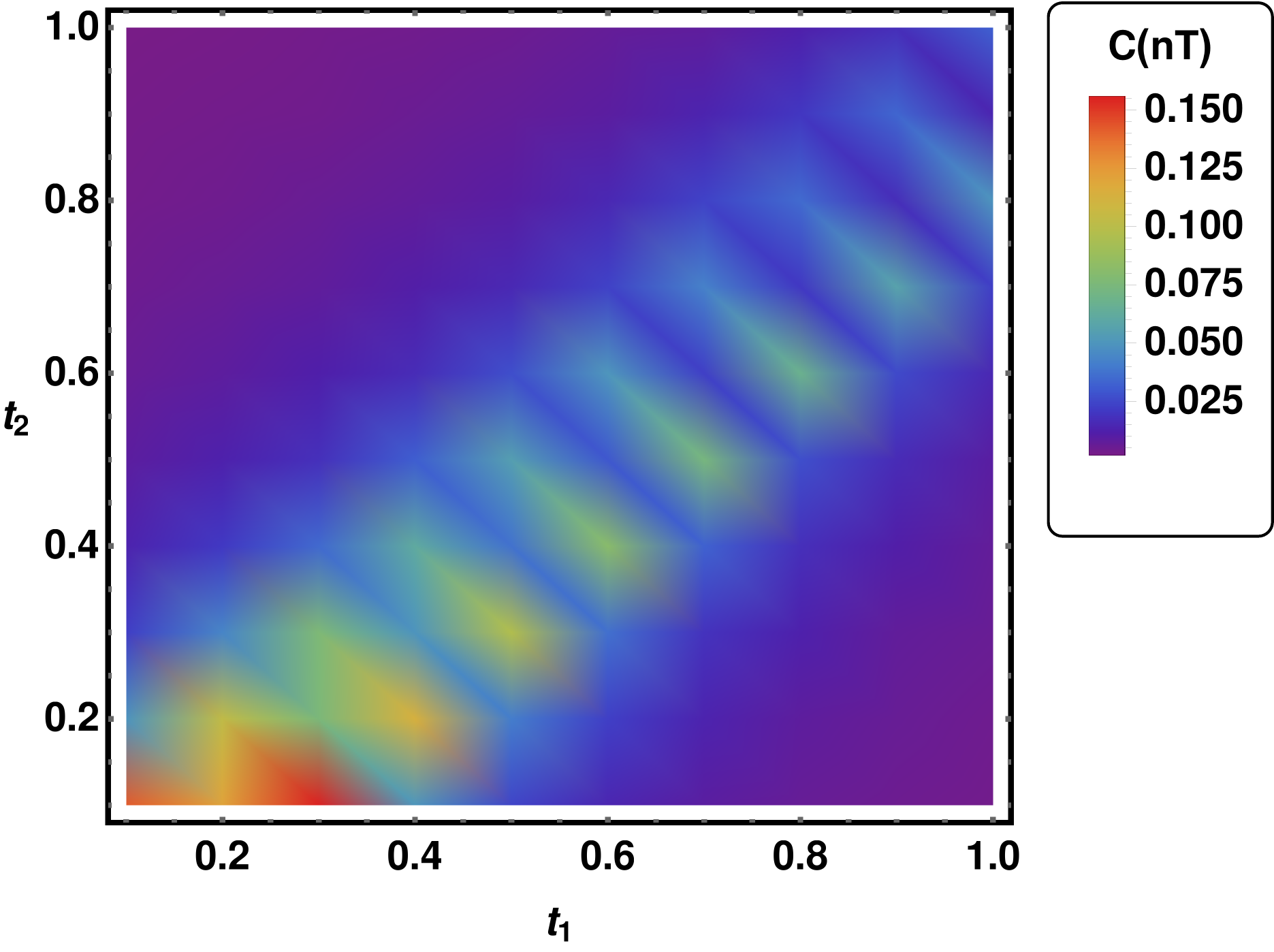}
		\label{sshpert1t2}}
	\caption{\small{Spread complexity in case of periodic variation of parameter for SSH model : (a)\,  Complexity with number of oscillations (b)\, Complexity with parameters }}
	\label{sshperiodic}
\end{figure}

\subsection{Complexity evolution with periodically driven field}

In presence of a periodically driven field  we take $t_{1}=t_{1}-v(t)$ and $t_{2}=t_{2}+v(t)$, with $v(t)=\delta \cos (\omega t)$. 
The Return amplitude is once again of the same form as the one in Eq. \eqref{return1/2}, with 
 \begin{equation}
 \gamma(t)=\arctan\Big(\frac{-|(t_{2}+v(t))\sin k|}{t_{1}-v(t)-(t_{2}+v(t))\cos k}\Big)~,    
 \end{equation}
 and
 \begin{equation}
\epsilon (t)=2\int_{0}^{t}\sqrt{[t_{1}-v(t')-(t_{2}+v(t'))\cos k]^{2}+[(t_{2}+v(t'))\sin k}]^{2}dt'~.     
 \end{equation}
With  periodic boundary conditions, in the continuum limit, the SC is given by 
\begin{equation}
    \mathcal{C}(nT)=\frac{1}{\pi}\int_{0}^{\pi}(1-|\mathcal{S}|^{2})~.
\end{equation}
In Fig. \ref{sshperiodic}, we have plotted SC in the presence of the periodically driven field. 
 Fig. \ref{sshoscn} shows  SC evolution  with respect to  a number of oscillations $n$. 
 Parameter values used are for blue curve: $\delta=0.1$, $\omega=\frac{2\pi}{T}$,  
 $T=1000$,  $t_1=0.5=t_2$,
  green curve: $\delta=0.1$, $\omega=\frac{2\pi}{T}$, 
 $T=1000$, $t_1=1$ , $t_2=0.2$, and the red curve:$\delta=0.1$, $\omega=\frac{2\pi}{T}$, $T=1000$, $t_1=0.2$, $t_2=1$.
In Fig. \ref{sshpert1t2}, complexity is plotted against  $t_1$ and $t_2$. Parameters for the plots: $\delta=0.1$, $\omega=\frac{2\pi}{T}$, $T=1000$, $n=40$. The non-analytical nature of the SC is easily seen around $t_{1}\approx t_{2}$ critical points.

\section{Statistics of the work done in quench }
In this section, we study the relationship between the SC evolution under 
different quench protocol  discussed so far in this paper, and an alternative way of looking at the quantum quench process \cite{polko},
namely as a thermodynamic transformation.
Our focus will  be on the work done $W$ on the system by quenching \cite{zhang, zhou}. $W$ is a random variable, and because of its 
fluctuating nature \cite{talkner, kurchan, jarz}, it can be  represented by the probability distribution $P(W)$
\cite{silva,spyros} 
\begin{equation}
    P(W)=\sum_{n\geq 0}\delta(W-E(n)+E_{0}) |\braket{n|\psi_{o}}|^{2}~.
\end{equation}
where $\ket{n}$ are the eigenstates of the post-quench Hamiltonian $H$ with energy $E(n)$, and
 $\ket{\psi_{0}}$ is the ground state  of the pre-quench Hamiltonian $H_{0}$ with eigenvalue $E_{0}$.

We first consider the  characteristics  function (CF) defined as the Fourier transform of probability distribution function $P(W)$
\begin{equation}
 G(t)\equiv <e^{-i W t}> =\int_{-\infty}^{\infty}dW e^{-i W t} P(W)~.
 \label{gen1}
\end{equation}
It can also be written as 
\begin{equation}
    G(t)=\bra{\psi_{0}}e^{i H_{0} t} e^{-i H t}\ket{\psi_{0}}=e^{i E_{0} t} \bra{\psi_{0}}e^{-i H t}\ket{\psi_{0}}~.
\end{equation}
The overall phase factor in front of the CF is usually neglected.
Taylor expanding CF  around $t=0$, from Eq. (\ref{gen1}), we have 
\begin{equation}
    G(t)\equiv <e^{-i W t}>=1+(-i t)<W>+\frac{1}{2}(-i t)^{2}<W^{2}>+O(t^{3})~.
    \end{equation}
We therefore  conclude that $<W> = i \frac{d G(t)}{dt}|_{t=0}$ and $<W^{2}> = i^{2} \frac{d^{2} G(t)}{dt^{2}}|_{t=0}$.

From the definition of the moments given in \cite{balas}:
\begin{equation}
    \mu_{n}=\frac{d^{n}}{dt^{n}}S(t)\big|_{t=0}=\bra{\psi(0)}\frac{d^{n}}{dt^{n}}e^{i H t}\ket{\psi(0)}\big|_{t=0}=\bra{K_{0}}(iH)^{n}\ket{K_{0}}~,
\end{equation}
we have the expression for the first few Lanczos coefficients as
\begin{equation}
    \bra{K_{0}}(iH)\ket{K_{0}}=i a_{0}~,~~
    \bra{K_{0}}(iH)^{2}\ket{K_{0}}=-a_{0}^{2}-b_{1}^{2}~.
\end{equation}
By using the definitions of the CF  and the return amplitude, $G(t)=S(t)^{*}$, we further obtain
\begin{equation}
i \frac{d}{dt}G(t)\big|_{t=0}=a_{0}~, \qquad  i^{2}\frac{d^{2}}{dt^{2}}G(t)\big|_{t=0}=a_{0}^{2}+b_{1}^{2}~.
\end{equation}

Thus we can express the average  and variance of the distribution of the work done in a quench 
for single mode of momentum mode to be 
\begin{equation}
\left\langle W \right \rangle =a_{0}~, \qquad \left\langle(\Delta W)^{2}\right \rangle= \left\langle W^{2}\right \rangle-\left\langle W\right \rangle^{2}=b_{1}^{2}~.
\end{equation}
In the continuum limit,  the total average and the variance of the work done can be written as  
\begin{equation}
\left\langle W \right \rangle =\frac{1}{2\pi}\int_{0}^{\pi}R_{f} \cos(\phi_{f}-\phi_{i})dk~,
\qquad \left\langle (\Delta W)^{2} \right \rangle =\frac{1}{2\pi}\int_{0}^{\pi}R^{2}_{f} \sin^{2}(\phi_{f}-\phi_{i}) dk~.
\end{equation}
For quench in the free fermion model considered above,  these two quantities can be expressed as 
\begin{equation}
\left\langle W \right \rangle= \frac{1}{2\pi}\int_{0}^{\pi}\frac{(R_{3f} R_{3i}+|R_{2f}||R_{2i}|)}{R_{i}}dk~,~\text{and}~~
\left\langle (\Delta W)^{2} \right \rangle=\frac{1}{2\pi}\int_{0}^{\pi}\frac{(|R_{2f}|R_{3i}-R_{3f}|R_{2i}|)^{2}}{R^{2}_{i}}~.
\label{w3}
\end{equation}
From these expressions, we see that the singularities in these quantities  depend on the initial values
of the parameters.

These expressions of work done and variance is the same in all three models we considered in this paper,
except the expressions  of $R_{3i}$, $R_{3f}$, $R_{2i}$, $R_{2f}$ which are determined by the 
the model under  consideration, and are functions of the parameters of particular models (with  the subscript $i$ and $f$
standing for the values of these parameters before and after quench).
E.g.  for three spin interaction transverse  Ising model
considered in sec \ref{3spin}, these are given in Eq. \eqref{para3}. On the other hand, for the spin 1/2 XY model 
considered in sec \ref{xyspin}, the expressions for these functions are given in Eq. \eqref{xydef}.
Similarly, for the SSH model considered in the previous section, see Eq. \eqref{sshdef} for the expression of these 
functions in terms of the parameters of this model.

\begin{figure}[ht!]
	\centering
	\subfigure[]{
		\includegraphics[width=0.3\textwidth]{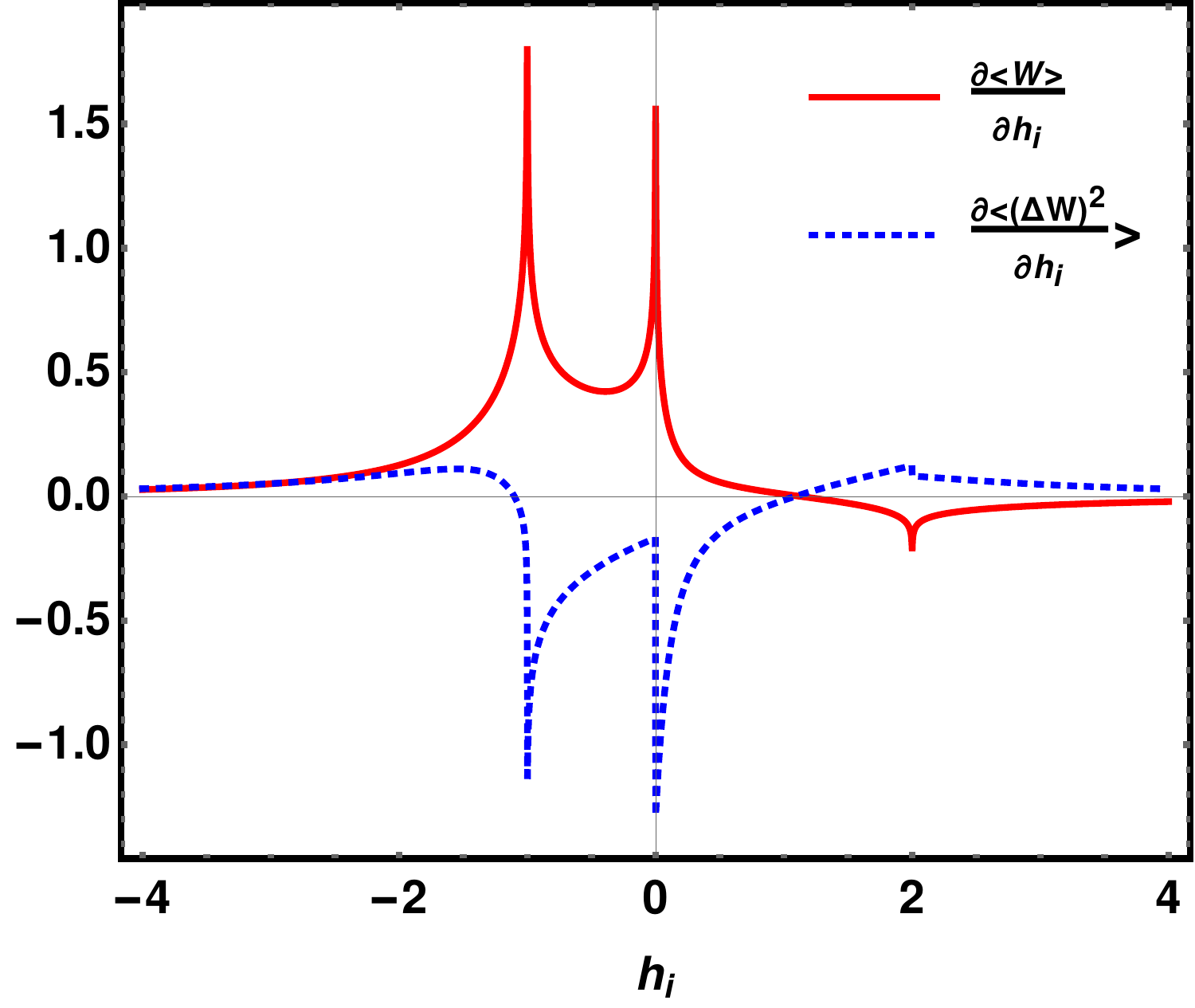}
		\label{fig: work3}}
	\centering
	\subfigure[]{
		\includegraphics[width=0.3\textwidth]{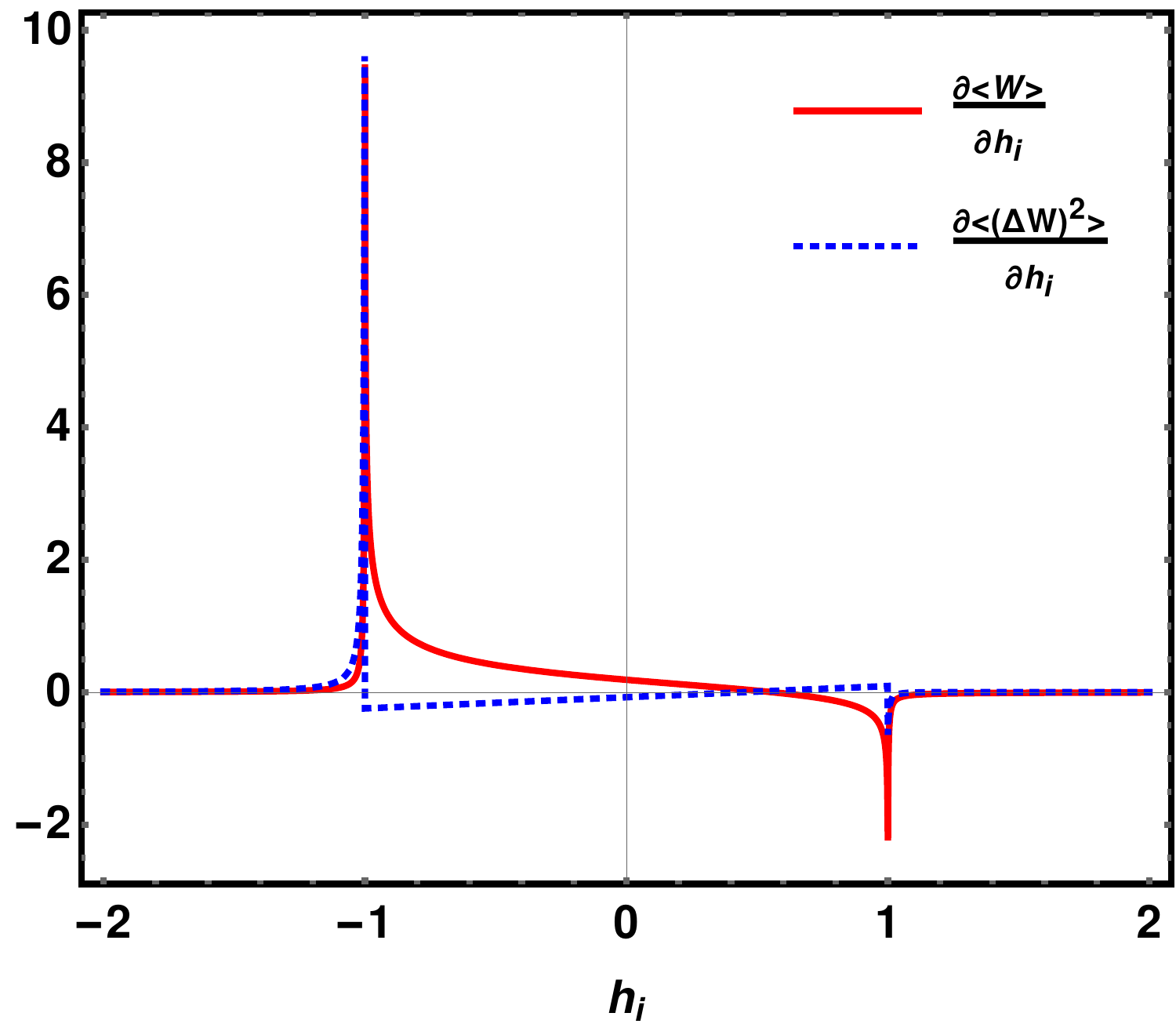}
		\label{fig:work2}}
	\centering
	\subfigure[]{
		\includegraphics[width=0.3\textwidth]{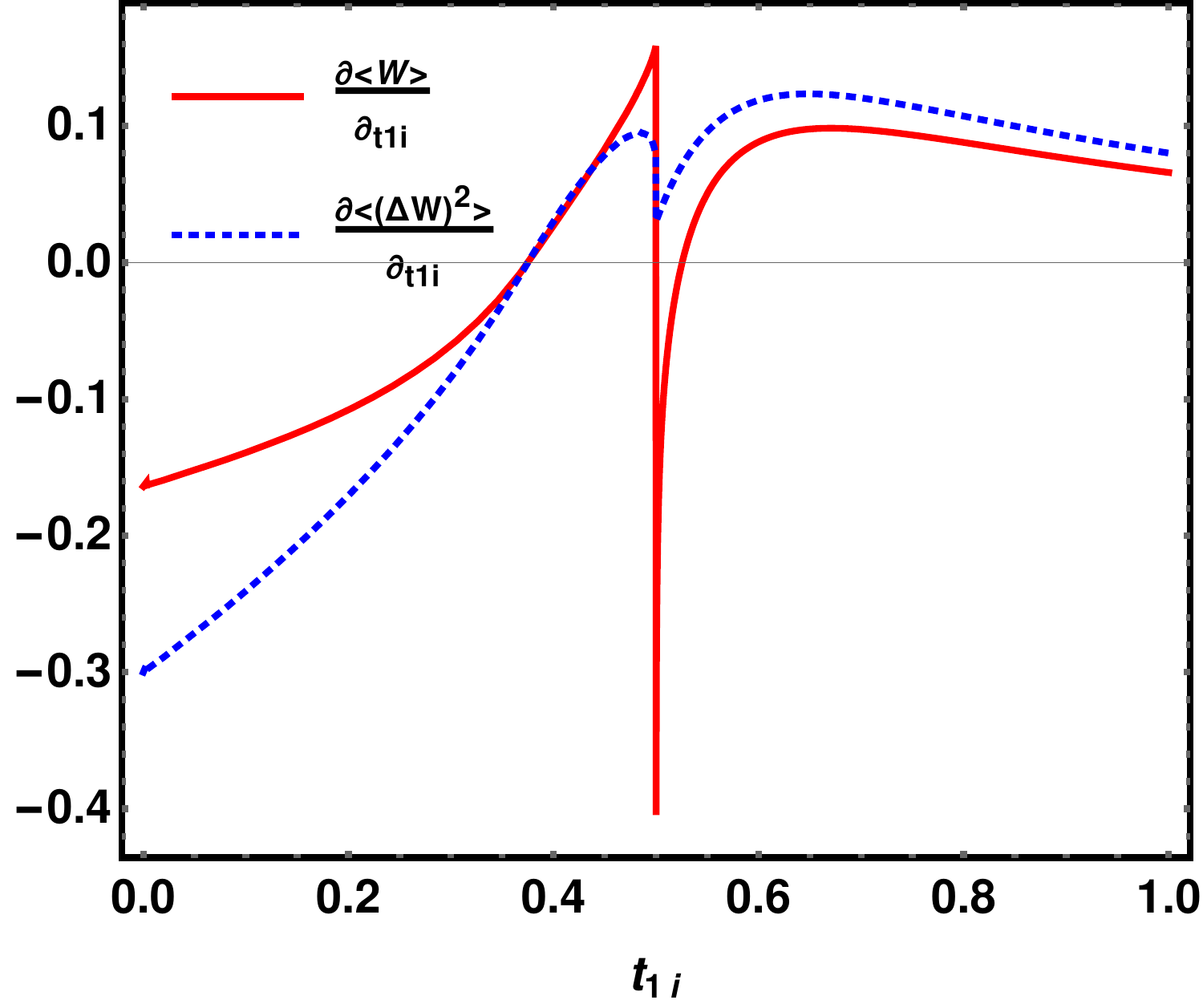}
		\label{fig:sshwva}}
	\caption{\small{Average and the variance of the statistics of work done : (a)\, Three spin interaction Ising model (b)\, XY spin chain model, and (c)\, SSH model.}}
	\label{fig:wo23spin}
\end{figure}

The  integral in the expression for the work done  is difficult  to evaluate analytically  in the three spin and XY model case, 
but the integral in the variance can be solved by using contour integration.
For simplicity,  here we present the results  for the SSH model only.  The work done on the SSH model can be expressed as 
\begin{equation}
   \left\langle  W \right \rangle = \frac{-1}{2\pi}\int_{0}^{\pi}R_{f}(\cos\phi_{f}\cos\phi_{i}
   +\sin\phi_{f}\sin\phi_{i}) dk~.
   \label{wossh}
\end{equation}
This integral can be performed analytically, and the final expression  can be written in terms of the elliptic functions
as 
 \begin{equation}
\left\langle  W \right \rangle= \frac{-1}{2\pi}\frac{1}{t_{1i}t_{2i}}\Bigg[(t_{1i}+t_{2i})(t_{1i}t_{2f}+t_{1f}t_{2i})\textit{F}_{1}
\Big[\frac{4t_{1i}t_{2i}}{(t_{1i}+t_{2i})^{2}}\Big]-(t_{1i}-t_{2i})(t_{1i}t_{2f}-t_{1f}t_{2i}) \textit{F}_{2}\Big[\frac{4t_{1i}t_{2i}}{(t_{1i}+t_{2i})^{2}}\Big]~.\label{w1ssh}
\Bigg]
\end{equation}
where $F_{1}$ and $F_{2}$ are elliptic integrals. Note that the above expression is valid only away from the critical point since Eq. (\ref{w1ssh})
 is not defined at ($t_{1i}=t_{2i}$).

Similarly, the variance of the work done can be expressed as 
 \begin{equation}
  \left\langle (\Delta W)^{2} \right \rangle=\begin{cases}
    \frac{(t_{1i}t_{2f}-t_{1f}t_{2i})^{2}}{4 t^{2}_{2i}}& \text{$|t_{2i}|>|t_{1i}|$\quad Topological phase}\\
    \frac{(t_{1i}t_{2f}-t_{1f}t_{2i})^{2}}{4 t^{2}_{1i}} ,& \text{$|t_{2i}|<|t_{1i}|$\quad Non Topological Phase}\\
  \end{cases}\label{varssh1}
\end{equation}
The above equation is continuous at $t_{1i}=t_{2i}$, but its derivative reflects non-analytical behaviour around the critical point.

We plot  the derivative of work done and its variance in Fig. \ref{fig:wo23spin} for three 
different free fermionic models considered. These quantities exhibit a non-analytical nature close to the critical points of the examined models. At $h=\pm 1$ for the XY model, at $h=J_3\pm1$, $h=-J_3$ for the three spin interaction model, and at $t_1=t_2$ for the SSH model, it shows non-analyticity. Parameters used  in these plots are,  Fig. \ref{fig: work3}: $J_{3f}=0.5$, $J_{3i}=1$, $h_{f}=1$, Fig. \ref{fig:work2}: $\gamma_{f}=0.5$, $\gamma_{i}=0.1$, $h_{f}=0.6$, and  Fig. \ref{fig:sshwva}: $t_{2f}=0.8$, $t_{2i}=0.5$, $t_{1f}=0.6.$

\section{Conclusions and discussions}
Our first motivation behind this paper was to see the qualitative difference in SC by considering three different types of interactions in spin chain models, the first with a two-spin interaction term, the second with a three-spin interaction term, and the third with dimerized interaction (SSH model). Because all three models are free fermion models, the evolution of SC is almost the same in each scenario.

The second motivation was to study SC as a probe to examine the quantum and topological phase transitions.
In essence, we have found that the derivative of the SC has a non-analytical behaviour at the critical point for all three models. We study the problem under single and multiple (Sudden) 
 quenches, where there is no sign of discontinuity near the critical point. Still, they attain steady state value for 
 a long time, as mentioned in \cite{captua}. We found that complexity reaches a steady state value earlier when the 
 the final value of the parameter is close to a critical point than when the parameter's initial value is at critical points; this behaviour applies to both single and multiple quench scenarios. In the time-dependent oscillating parameter case, when we plot the complexity with respect to the parameter (in the case of three spin and XY spin, it is $h_c$, while for SSH, it is $t_1$, $t_2$), there are peaks at the critical value of the parameter. On the other hand, when we examine the complexity with respect to a number of oscillations, first, it reaches the maximum value. Then it oscillates around an average value without any sign of discontinuity at the critical point. 
 
 Furthermore, we have also calculated the statistics of the  work done under quench and explained their relationship with the Lanczos coefficients. The derivative of average work done and variance with respect to the parameter displayed non-analytical behaviour around critical points in all three models.

In this paper, we have confined our attention to the cases where the model can be mapped to a free-fermion model, which in turn made an analytical approach possible. It will be interesting to study the dynamics of SC in the more generic model, where this kind of mapping is not  possible, which we left for future work.

\section{Acknowledgements}
We thank Prof. Tapobrata Sarkar, Kunal Pal  and Kuntal Pal for their guidance and discussions. We are also thankful to Pritam Banerjee for his help and suggestions. We appreciate Pratik Nandy and Jacques H.H. Perk's concern and important suggestions.

\appendix

\section{ Details of three spin interaction Ising model}\label{3spinapp}
For three spin case in Eq. (\ref{three hamilt}),
in JW transformation, Pauli matrices can be written in terms of the Fermionic operator as 
 \begin{equation}
     c_{i}=\sigma^{-}_{i}\exp\Big(-i\pi\sum_{j=1}^{i-1}\sigma^{+}_{j}\sigma^{-}_{j}\Big)~,~~
      \sigma^{z}_{i}=2 c^{\dagger}_{i}c_{i}-1~,
\end{equation}
where $\sigma^{+}=(\sigma^{x}+i\sigma^{y})/2$, $\sigma^{-}=(\sigma^{x}-i\sigma^{y})/2$, and they satisfy the following  
      standard anticommutation relations
     \[
     [c^{\dagger}_{i} c_{j}]_{+}=\delta_{ij},~~\text{and}~~
     [c^{\dagger}_{i},c^{\dagger}_{j}]_{+}=[c_{i},c_{j}]_{+}=0~.
     \]
Now using the  Fourier transformation mentioned in the main text, we can write the Hamiltonian as
     \begin{equation}
         H=-\sum_{k>0}(h+\cos{k}-J_3\cos{2k})(c^{\dagger}_{k}c_{k}+c^{\dagger}_{-k}c_{-k})
         +i(\sin{k}-J_3\sin{2k})(c^{\dagger}_{k}c^{\dagger}_{-k}+c_{k}c_{-k})
         \label{three fourier a}~.
     \end{equation}
This  can be rewritten  in the following form
     \begin{equation}
         H=-\sum_{k}\Big[\mathbf{I}(h+\cos{k}-J_3\cos{2k})+\mathbf{\Phi}_{k}^{\dagger}\vec{R}(k).\vec{\sigma}\mathbf{\Phi}_{k}
         \Big]~,
     \end{equation}
 where $\mathbf{I}$ is identity matrix, $\mathbf{\Phi}_{k}=\begin{bmatrix}
    c_k \\
    c_{-k}^{\dagger} \\
\end{bmatrix}$, $\vec{R}=\begin{bmatrix}
    0 \\
    J_3 \sin 2k-\sin k \\
    h+\cos k-J_3\cos 2k
    \\
   \end{bmatrix} $, and $\vec{\sigma}$ = $(\sigma_{x},\sigma_{y},\sigma_{z})$.
Here we consider periodic boundary conditions, so that  $k=\frac{2n\pi}{L}$,  where $n=,1,2,3\cdots \frac{N}{2}$, assuming $N$ is even. Furthermore, we can  diagonalize the above Hamiltonian  by using a  Bogoliubov transformation of the form
\begin{align}\label{bogoliubov}
 \eta_{k}=-\cos{(\frac{\phi}{2})}c^{\dagger}_{-k}+i\sin{(\frac{\phi}{2})}c_{k}\\
\eta^{\dagger}_{-k}=i\cos{(\frac{\phi}{2})}c_{k}+\sin{(\frac{\phi}{2})}c^{\dagger}_{-k}~,
\end{align}
where we have defined
 \[
\cos{\phi}=\frac{R_3}{R};\qquad
R_3=h+\cos{k}-J_3\cos{2k}~,\qquad R_2=J_3\sin{2k}-\sin{k}~,\qquad R=\sqrt{R_2^{2}+R_3^{2}}=|\vec{R}(k)|~.
\]
The diagonal form of the Hamiltonian can be written as 
\begin{equation}
    H=\sum_{k>0}R(k)[-\eta_{-k}\eta_{-k}^{\dagger}+\eta_{k}^{\dagger}\eta_{k}]~.
\end{equation}
The ground state is defined by the condition $\eta_{k}\ket{\psi_{gs}}=0$ can be written as 
\begin{equation}\label{3ground}
\ket{\psi_{gs}}=\prod_{k>0}\sin{(\frac{\phi}{2})}\ket{0,0}-i\cos{(\frac{\phi}{2})}\ket{1,1}
\end{equation}

Now we describe how the above Hamiltonian can be written as an element of the  $su(2)$ algebra. 
$su(2)$  algebra is defined by the following commutation relations
 \[
 [J_{z},J_{\pm}]=\pm J_{\pm}, \quad [J_{+},J_{-}]=2 J_{z}, \quad [J_{\pm},I]=0, \quad [J_{z},I]=0~.
 \]
 We have to find out the operator in our problem which satisfies the above relations; then we represent them by $J_{z}$, $J_{+}$, $J_{-}$. 
To this end, we first define 
\begin{equation}\label{su2_generators}
J^{k}_{0}=\frac{1}{2}(c^{\dagger}_{k}c_{k}-c_{-k}c^{\dagger}_{-k})~,~J^{k}_{+}=c^{\dagger}_{k}c^{\dagger}_{-k}~,~~
J^{k}_{-}=c_{-k}c_{k}~,
\end{equation}
so that the form of the above Hamiltonian in terms of these operators can  be written as 
 \[
 H=\sum_{k>0}R_{1}I^{k}+R_{1}J_{z}^{+k}+ \frac{i}{2}R_2 J_{+}^{+k}- \frac{i}{2}R_2 (J_{-}^{+k})~,
 \]
here $R_1$, and $R_2$ can be replaced by $R\cos{\phi}$, $R\sin{\phi}$, respectively,  where
 $R=\sqrt{R_{1}^{2}+R_2^{2}}$. 
 \\

\section{Details of the XY spin chain}\label{xyapp}

In this appendix, we show that the Hamiltonian of the XY spin chain can be written as an element of the $su(2)$ Lie algebra
and obtain the ground state of the model.
The relation between the  $su(2)$ generators and the creation and annihilation operators are the same as those  in 
 Eq. \eqref{su2_generators}. Using these the Hamiltonian in  Eq. (\ref{xyjw}) can be rewritten as 
\begin{equation}
H=\sum_{k>0}\Big[-2(h+\cos{k})J^{k}_{0}+i\gamma\sin{k}J^{k}_{+}-i\gamma\sin{k}J^{k}_{-}\Big]
  =\sum_{k>0}\Big[-2(R \cos\phi)J^{k}_{0}+i R \sin\phi J^{k}_{+}-i R \sin\phi J^{k}_{-}\Big]~,
\label{xysu12R}
\end{equation}
where we have defined 
$R(k)=\sqrt{R_2^{2}+R_3^{2}}$, with $R_2=\gamma \sin{k}$, $R_3=h+\cos{k}$, $\phi=\tan^{-1}\frac{|R_2|}{R_3}$.
By using the Bogoliubov transformations of the form written in Eq. \eqref{bogoliubov}, with 
$\phi=\cos^{-1}\Big[\frac{h+\cos k}{\gamma \sin k}\Big]$ we diagonalize  the Hamiltonian of Eq. (\ref{xysu12R}), so that
the final form of the Hamiltonian is 
\begin{equation}
H=\sum_{k>0}R(k)\Big[\eta^{\dagger}_{k}\eta_{k}+\eta^{\dagger}_{-k}\eta_{-k}-1\Big]~.
\end{equation}

The form for the ground state is defined by the condition $\eta_{k}\ket{\psi_{g}}=0$ is the same as in Eq. \eqref{3ground},
and it can be written as a $SU(2)$ coherent state of the form
\begin{equation}
\ket{\psi_{g}}=\prod_{k>0}\sin\frac{\phi_{i}}{2}e^{-i\cot\frac{\phi_{i}}{2}J_{+}^{k}}\ket{\frac{1}{2},\frac{-1}{2}}_{k}~.
\end{equation}

\section{Duality Transformation}\label{dutr}
This section will explore the relationship between the XY-spin chain model and the three-spin interaction TIM under the duality transformation. For details, see \cite {jjh,jjh1}. 
\[
A_{n}=\sum_{j=1}^{N}\sigma^{x}_{j}\left(\prod^{j+n-1}_{k=j+1}\sigma^{z}_{k}\right)\sigma^{x}_{j+n}
\]
\[
A_{-n}=\sum_{j=1}^{N}\sigma^{y}_{j}\left(\prod^{j+n-1}_{k=j+1}\sigma^{z}_{k}\right)\sigma^{y}_{j+n}
\]
\[
A_{0}=-\sum_{j=1}^{N}\sigma^{z}_{j} ,
\quad
A_{1}=\sum_{j=1}^{N}\sigma^{x}_{j}\sigma^{x}_{j+1}
\quad
A_{-1}=\sum_{j=1}^{N}\sigma^{y}_{j}\sigma^{y}_{j+1}
\quad
A_{2}=\sum_{j=1}^{N}\sigma^{x}_{j}\sigma^{z}_{j+1}\sigma^{x}_{j+2}
\]
\[
\tau^{z}_{n}=\sigma^{x}_{n-1}\sigma^{x}_{n} ,\quad
\tau^{x}_{n}=\prod^{n-1}_{k=1}\sigma^{z}_{k}
\]
where $\tau^{\alpha}_{n}$ dual quantum spins, satisfying Pauli spin algebra located at the centres of the bonds of the original lattice
\begin{equation}
\mathcal{H}=-\frac{1}{2}\Big(\sum_{i}\sigma^{z}_{i}\Big[h+J_3\sigma^{x}_{i-1}\sigma^{x}_{i+1}\Big]
+J_{x}\sum_{i}\sigma^{x}_{i}\sigma^{x}_{i+1}\Big)\label{threedu}
\end{equation}
Above Hamiltonian consist of $A_{0}$ , $A_{1}$ and $A_{2}$, where as XY-spin chain Hamiltonian Eq.(\ref{hamit xy}) consist of $A_{0}$ , $A_{1}$ , $A_{-1}$.
By using dual representation, the Hamiltonian Eq.(\ref{threedu}) can be written in XY spin chain form
\begin{equation}
\mathcal{H}=-\frac{1}{2}\Big(\sum_{n}h\tau^{x}_{n}\tau^{x}_{n+1}-J_3\tau^{y}_{n}\tau^{y}_{n+1}
+J_{x}\sum_{n}\tau^{z}_{n}\Big)
\end{equation}
XY-spin chain model in general form 
\begin{equation}
    H=-\sum_{i}\big[J'_{x}\sigma^{x}_{i}\sigma^{x}_{i+1}+J_{y}\sigma^{y}_{i}\sigma^{y}_{i+1}+h'\sigma^{z}_{i}]\label{xyd}
\end{equation}
Eq.(\ref{threedu}) can map into Eq.(\ref{xyd}) with $\frac{h'}{J'_{x}-J_{y}}=\frac{J_{x}}{h+J_{3}}$ and $\frac{h'}{J'_{x}+J_{y}}=\frac{J_{x}}{h-J_{3}}$
\section{Auxillary Equation for time-dependent case}  \label{timede1}
In this appendix, we will find out the relation of $\gamma(t)$ and $\beta(t)$ defined in Eq. (\ref{rt1}). For details, see Refs. \cite{lai,Maam}.
Comparing Eqs. (\ref{SU2_three2}), (\ref{xysu2R}), and (\ref{sshsu2}), we  find out that the Hamiltonian in our problem has 
the following form  (with $\phi(t)=\frac{\pi}{2}$)
\[
H=f(t)J_{0}^{k}+g(t)[J_{+}^{k}e^{i\phi(t)}+J_{-}^{k} e^{-i\phi(t)}]~.
\]
where $f(t)$, $g(t)$, and $\phi(t)$ are real functions of time.
The Schrodinger equation governs the time evolution of the quantum state.
\[
i\frac{d}{dt}\ket{\psi(t)}=H(t)\ket{\psi(t)}~.
\]
Next, we define the invariant operator $\mathcal{I}(t)$, which satisfies the following conditions
\begin{equation}
 \frac{d\mathcal{I}}{dt}\equiv i\frac{\partial\mathcal{I}}{\partial t}+[\mathcal{I},H]=0 ~,~~\text{and}~~
 \mathcal{I}^{\dagger}=\mathcal{I}~.
 \label{invar}
\end{equation}
Using the above conditions for the invariant operator, we see that it satisfies the equation.
\[
i\frac{d}{dt}(\mathcal{I}\ket{\psi(t)})=H(t)(\mathcal{I}\ket{\psi(t)})~.
\]
Therefore if $\ket{\psi(t)}$ satisfy Schrodinger equation so does $\mathcal{I}\ket{\psi(t)}$.
For our purposes, we define the invariant operator as
\begin{equation}
\mathcal{I}(t)=\hat{R}(t)\hat{J_{0}}\hat{R}^{\dagger}(t)\label{it} ~,
\end{equation}
where 
\begin{equation}
\hat{R}(t)=\exp\Big[\frac{\gamma(t)}{2}\Big(\hat{J}_{+} e^{-i\beta(t)}-\hat{J}_{-} e^{i\beta(t)}\Big)\Big]~.
\label{rt}    
\end{equation}
Time-dependent parameter $\gamma(t)$ and $\beta(t)$ are related to the functions  $f(t)$ and $g(t)$ 
appearing in the Hamiltonian through
\begin{equation}
\dot{\gamma}=2g(t)\sin{(\phi+\beta)}~,
\qquad
\big(\dot{\beta}-f(t)\big)\sin{\gamma}=2 g(t)\cos{\gamma}\cos{(\phi+\beta)} ~.   
\end{equation}
We can derive the above equations by using  Eqs. 
 (\ref{invar}), (\ref{rt}), and (\ref{it}) and the following relations
\[
\hat{R}^{\dagger}(t)\hat{J}_{+}\hat{R}(t)=\hat{J}_{+}\cos^{2}\frac{\gamma}{2}
-\hat{J}_{-}e^{2i\beta}\sin^{2}\frac{\gamma}{2}-\hat{J}_{0}e^{i\beta}\sin\gamma~,
\]
\[
\hat{R}^{\dagger}(t)\hat{J}_{-}\hat{R}(t)=\hat{J}_{-}\cos^{2}\frac{\gamma}{2}
-\hat{J}_{+}e^{-2i\beta}\sin^{2}\frac{\gamma}{2}-\hat{J}_{0}e^{-i\beta}\sin\gamma~,
\]
\[
\hat{R}^{\dagger}(t)\hat{J}_{0}\hat{R}(t)=\hat{J}_{0}\cos\gamma+\frac{1}{2}(\hat{J}_{+}e^{-i\beta}
+\hat{J}_{-}e^{i\beta})\sin\gamma~,
\]
as well as
\[
\hat{R}^{\dagger}(t)\Big[i\frac{\partial \hat{R}(t)}{\partial t}\Big]=-2\hat{J}_{0}\dot{\beta}\sin^{2}\frac{\gamma}{2}+\hat{J}_{+}e^{-i\beta}
\Big(i\frac{\dot{\gamma}}{2}+\frac{\dot{\beta}}{2}\sin\gamma\Big)+{J}_{-}e^{i\beta}\Big(-i\frac{\dot{\gamma}}{2}
+\frac{\dot{\beta}}{2}\sin\gamma\Big)~,
\]
To solve the differentiation of an operator by a parameter, we use the relation given below (where $\lambda$ represents the parameter on which operator $A$ depends)
\[
\frac{\partial \exp{\hat{A}}}{\partial \lambda}=\int_{0}^{1} ds\exp{(1-s)\hat{A}}\frac{\partial{\hat{A}}}{\partial \lambda}\exp{s\hat{A}}~.
\]
Since
$\phi=\frac{\pi}{2}$
and by symmetry, we choose
$\beta=-\phi$.
This led to the following equations.
\[
\dot{\gamma}\approx 0~~\text{and}~~-f(t)\sin{\gamma}=2 g(t)\cos{\gamma}~.
\]
From the second equation, we obtain 
 \[
\gamma(t)=\tan^{-1}\Big[-\frac{2g(t)}{f(t)}\Big]~.
\]
To satisfy the condition $\dot\gamma(t)\approx 0$ 
we have to take  $\omega\approx \textit{small}$ (adiabatic approximation, the variation in $\gamma(t)$ is negligible).

\section{Time Dependent state under periodic variation of parameter}\label{timede2}

In this section, we find the explicit expression of time evolved state under periodic variation of parameters.
Let $\ket{n}$ be the eigenstate of $J_{0}$ with eigenvalue $j$ 
\[
J_{0}\ket{n}=j\ket{n}
\]
Eigenstate of $\mathcal{I}(t)$
\[
\mathcal{I}(t)\ket{n,t}=j\ket{n,t}
\qquad
\ket{n,t}=\hat{R}(t)\ket{n}
\]
According  to the LR theory \cite{Lewis}
the eigenvectors of $\mathcal{I}(t)$ are related to the general state by time-dependent gauge transformation
\[
\ket{\psi(t)}=\sum_{n}c_{n} e^{i\alpha_{n}(t)} \ket{n,t}~,
\]
where $c_{n}$ are time-independent coefficients, and the time-dependent phase factor $\alpha_{n}(t)$ can be written as 
\begin{equation}
    \alpha_{n}(t)=\int_{0}^{t}dt'\bra{n,t'}i\frac{\partial}{\partial t'}-\hat{H}(t')\ket{n,t'}
    =\int_{0}^{t}dt'\bra{n}[\hat{R}^{\dagger}(t')i\frac{\partial}{\partial t'}\hat{R}(t')-\hat{R}^{\dagger}(t')\hat{H}(t')\hat{R}(t')]\ket{n}~.
\end{equation}
which can be simplified to:
\[
\alpha_{n}(t)=-j\int_{0}^{t}dt'\Big[f(t')+2(\dot{\beta}-f(t))\sin^{2}{\frac{\gamma}{2}}-2 g(t)\sin{\gamma}\cos{(\phi+\beta)}
\Big]
\]
and 
\[
\epsilon (t)=\int_{0}^{t}dt'[f(t')+2(\dot{\beta}-f(t))\sin^{2}{\frac{\gamma}{2}}-2 g(t)\sin{\gamma}\cos{(\phi+\beta)}]~.
\]
At $t=0$, we have $\ket{\psi(0)}=\hat{R}(0)\sum_{n}c_{n}\ket{n}$. In our case, we take $\ket{\psi(0)}$ as the ground state of the initial Hamiltonian, i.e. the one given in Eq. \eqref{tpstate}.
Now by using the BCH formula, we obtain  
 \[
 \hat{R}^{\dagger}(0)=\exp(z_{0}J_{+})\exp(\log(1+z_{0}\bar{z}_{0})J_{0})\exp(-\bar{z_{0}}J_{-})~,~~\text{with}~~
 z_{0}=-i\tan(\frac{\gamma(0)}{2})~,~~\text{and}~~
 \]
 \[
\hat{R}(t)=\exp(z_{t}J_{+})\exp(\log(1+z_{t}\bar{z}_{t})J_{0})\exp(-\bar{z_{t}}J_{-})~,~~\text{with}~~
 z_{t}=i\tan(\frac{\gamma(t)}{2})~,~~\text{and}~~
 \]
 \[
\hat{R}_{i}=\exp(z_{i}J_{+})\exp(\log(1+z_{i}\bar{z}_{i})J_{0})\exp(-\bar{z_{i}}J_{-})~,~~\text{with}~~
 z_{i}=-i\cot(\frac{\phi_{i}}{2})~.
 \]
 \begin{equation}
 \begin{split}
  \ket{\psi(t)}=\prod_{k>0}\cos^{2j}{\frac{\gamma(0)}{2}} \sin^{2j}(\frac{\phi_{i}}{2})\cos^{2j}{\frac{\gamma(t)}{2}} e^{-i j\epsilon(t)}(1+z_{i}z_{0})^{2j}  (1+z_{t} \alpha e^{-i\epsilon(t)})^{2j} \\   
   \exp\Big[[z_{t}+\frac{\alpha \sec^{2}{\frac{\gamma(t)}{2}} e^{-i\epsilon(t)}}{1+\alpha z_{t} e^{-i\epsilon(t)}}]J_{+}^{k}]\Big]\ket{j,-j}_{k}  
 \end{split}
 \end{equation}
where
$\alpha= z_0+\frac{z_i\sec^2[\frac{\gamma(0)}{2}]}{1+z_i z_0}$.
The return amplitude here can be written as
 \[
 \mathcal{S}(t)=\sum_{n=0}^{2j}(-z_{i})^{n}\frac{(2j)!}{(2j-n)!n!}\Big[z_{t}+\frac{\alpha \sec^{2}{\frac{\gamma(t)}{2}} e^{-i\epsilon(t)}}{1+\alpha z_{t} e^{-i\epsilon(t)}}\Big]^{n} A(t)~,
 \]
where 
 \[
 A(t)=
 \cos^{2j}\Big[{\frac{\gamma(0)}{2}}\Big]\sin^{4j}\Big[{\frac{\phi_{i}}{2}}\Big]\cos^{2j}\Big[{\frac{\gamma(t)}{2}}\Big]
 e^{-i j\epsilon(t)}\big(1+z_{i}z_{0}\big)^{2j} 
 \big(1+z_{t} \alpha  e^{-i\epsilon(t)}\big)^{2j}~.
 \]
When $j=\frac{1}{2}$, we get back the expression for the return amplitude provided in Eq. \eqref{return1/2}.

For periodic condition $\gamma(nT)=\gamma(0)=\gamma$
 \begin{equation}
 \begin{split}
\ket{\psi(nT)}=\prod_{k>0}\Big[[\cos(\frac{\epsilon(nT)}{2}) \sin(\frac{\phi_{i}}{2})-i\sin(\frac{\epsilon(nT)}{2})\sin(\gamma-\frac{\phi_{i}}{2})]\ket{\frac{1}{2},\frac{-1}{2}}_{k}+ \\    
[-\cos(\gamma-\frac{\phi_{i}}{2})\sin(\frac{\epsilon(nT)} {2})-i\cos(\frac{\epsilon(nT)}{2})\cos(\frac{\phi_{i}}{2})]\ket{\frac{1}{2},\frac{1}{2}}_{k}\Big]~.
 \end{split}
 \end{equation}

\end{document}